\documentclass[twocolumn]{aastex631}
\usepackage[utf8]{inputenc}
\usepackage{amsmath}

\newcommand{\hMpcinv}{h \,\rm Mpc^{-1}}
\newcommand{\kpara}{k_\parallel}
\newcommand{\kperp}{k_\perp}

\begin{document}

\title{The Impact of Beam Variations on Power Spectrum Estimation for 21\,cm Cosmology II:\\Mitigation of Foreground Systematics for HERA}

\correspondingauthor{Honggeun Kim}
\email{hgkim@mit.edu}

\author[0000-0001-5421-8927]{Honggeun Kim}
\affiliation{Department of Physics, Massachusetts Institute of Technology, Cambridge, MA, USA}
\affiliation{MIT Kavli Institute, Massachusetts Institute of Technology, Cambridge, MA, USA}

\author[0000-0002-8211-1892]{Nicholas S. Kern}
\affiliation{Department of Physics, Massachusetts Institute of Technology, Cambridge, MA, USA}
\affiliation{MIT Kavli Institute, Massachusetts Institute of Technology, Cambridge, MA, USA}

\author[0000-0002-4117-570X]{Jacqueline N. Hewitt}
\affiliation{Department of Physics, Massachusetts Institute of Technology, Cambridge, MA, USA}
\affiliation{MIT Kavli Institute, Massachusetts Institute of Technology, Cambridge, MA, USA}

\author[0000-0001-5122-9997]{Bang D. Nhan}
\affiliation{Central Development Laboratory, National Radio Astronomy Observatory, Charlottesville, VA, USA}

\author[0000-0003-3336-9958]{Joshua S. Dillon}
\affiliation{Department of Astronomy, University of California, Berkeley, CA, USA}

\author[0000-0001-8530-6989]{Eloy de~Lera~Acedo}
\affiliation{Cavendish Astrophysics, University of Cambridge, Cambridge, UK}

\author[0000-0001-7010-0937]{Scott B. C. Dynes}
\affiliation{MIT Kavli Institute, Massachusetts Institute of Technology, Cambridge, MA, USA}

\author[0000-0003-2560-8023]{Nivedita Mahesh}
\affiliation{Cahill Center for Astronomy and Astrophysics, California Institute of Technology, Pasadena, CA, USA}

\author[0000-0001-5300-3166]{Nicolas Fagnoni}
\affiliation{Cavendish Astrophysics, University of Cambridge, Cambridge, UK}

\author[0000-0003-3197-2294]{David R. DeBoer}
\affiliation{Department of Astronomy, University of California, Berkeley, CA, USA}

\date{May 2023}

\begin{abstract}
One key challenge in detecting 21\,cm cosmological signal at $z>6$ is to separate the cosmological signal from foreground emission. This can be studied in a power spectrum space where the foreground is confined to low delay (or equivalently, $k_\|$) modes whereas the cosmological signal can spread out to high delay modes. When there is a calibration error, however, chromaticity of gain errors propagates to the power spectrum estimate and contaminates the modes for cosmological detection. The Hydrogen Epoch of Reionization Array (HERA) employs a high-precision calibration scheme using redundancy in measurements. In this study, we focus on the gain errors induced by nonredundancies arising from feed offset relative to the HERA's 14-meter parabolic dish element, and investigate how to mitigate the chromatic gain errors using three different methods: restricting baseline lengths for calibration, smoothing the antenna gains, and applying a temporal filter prior to calibration. With 2~cm/2$^\circ$ perturbations for translation/tilting motions, a level achievable under normal HERA operating conditions, the combination of the baseline cut and temporal filtering indicates that the spurious gain feature due to nonredundancies is significantly reduced, and the power spectrum recovers the clean foreground-free region. We found that the mitigation technique works even for large feed motions but in order to keep a stable calibration process, the feed positions need to be constrained to 2~cm for translation motions and 2$^\circ$ for tilting offset relative to the dish's vertex.
\end{abstract}

\section{Introduction}

Understanding the formation of the first stars and galaxies in the universe is key to our broader picture of structure formation over cosmic time. This era, known as Cosmic Dawn, and the subsequent Epoch of Reionization (EoR), when these sources injected ionizing photons into the intergalactic medium (IGM), set the stage for the emergence of modern galaxies as we see them in the present universe; however, we currently have a poor understanding of the astrophysics of these eras. A powerful probe of Cosmic Dawn astrophysics is the 21\,cm hyperfine signal emanating from neutral hydrogen in the IGM at redshifts $z>6$. With the efforts to detect the cosmological signal, there are past, current and upcoming instruments including the Giant Metre Wave Radio Telescope \citep[GMRT;][]{Paciga2013}, the Murchison Widefield Array \citep[MWA;][]{Tingay2013, Dillon2014, Beardsley2016, Ewall-Wice2016, Barry2019, Trott2020}, the Donald C. Backer Precision Array for Probing the Epoch of Reionization \citep[PAPER;][]{Parsons2010, Cheng2018, Kolopanis2019}, the Low Frequency Array \citep[LOFAR;][]{vanHaarlem2013, Patil2017, Gehlot2019, Mertens2020}, the Hydrogen Epoch of Reionization Array \citep[HERA;][]{Dillon2016, DeBoer2017, Kern2022, Dillon2023}, and the Square Kilometre Array \citep[SKA;][]{Koopmans2015}.

A key challenge in making robust measurements of 21\,cm emission at the EoR are the bright synchrotron foreground emission coming from the foreground galaxy and from extragalactic sources, which are at least 4-5 orders of magnitude brighter. One of the ways to circumvent the foreground obstacle is to employ a 2D power spectrum approach that has a potential to separate the cosmological signals from the foregrounds. The spectrally smooth foreground emission, in principle, is confined to a certain range of the delay ($\tau$) or the line-of-sight cosmological modes ($\kpara$), a Fourier dual to the frequency. In contrast, the cosmological signals trace the evolution of the neutral hydrogen with fluctuations in a brightness temperature field along the redshift or frequency, and hence are distributed across a wide range of the delay modes. This provides foreground-free delay modes for the cosmological detection \citep{Datta2010, Vedantham2012, Morales2012, Morales2018, Parsons2012, Trott2012, Thyagarajan2013, Liu2014, Pober2014}.

In order to adopt the strategy described above, high-precision antenna gain calibration is required. Especially if there are frequency-dependent (chromatic) calibration gain errors larger than in a fraction of $10^{-5}$, the spectral artifacts from the gain solutions can significantly reduce the detectability of the 21\,cm cosmological signals. Redundant baseline calibration proposes a way of precise calibration that uses redundancy in measurements of baselines that sample the same Fourier mode \citep{Wieringa1992, Liu2010, Dillon2020}. If all antennas are identical, the calibration process should return exact gain solutions independent of our knowledge of the sky. However, if there are nonredundancies in the data, for example due to per-antenna perturbations, the redundant calibration can give rise to spurious chromaticity in gain solutions \citep{Byrne2019, Orosz2019, Choudhrui2021}. Furthermore, redundant calibration schemes have degenerate degrees of freedom that must be solved using an external sky model \citep{Liu2010, Zheng2017, Dillon2018, Byrne2019, Kern2020a}.

\citet[][hereafter, Paper~I]{Kim2022} numerically model the deformation of the primary beams for HERA according to different types of per-antenna feed perturbations, including vertical, horizontal, and tilting offsets. Such a beam deformation can introduce nonredundancies into measurements and cause chromatic gain errors from redundant calibration. As a result, the chromatic gain errors will give rise to the leakage of foreground modes into the cosmological modes, which are supposed to be foreground-free in the ideal redundant calibration.

\citet{Ewall-Wice2016} and \citet{Orosz2019} look primarily at the nonredundancy effect on gain errors with simplistic per-antenna primary beam deformations. They show that down-weighting the contribution of long baselines to the calibration can reduce the calibration bias and alleviate the foreground contamination in the 2D power spectrum. Motivated by these studies, we try to mitigate the spectral gain errors found in Paper~I with various mitigation methods. 

In this study, similar to Paper~I, we focus on the Phase~II system of HERA with zenith-pointing 14-m dish elements. We explore three different techniques of mitigating the chromatic gain errors. These are
\begin{itemize}
    \item \emph{Baseline cut}: we restrict the baseline lengths used for calibration, similar to \citet{Ewall-Wice2016} and \citet{Orosz2019}.
    \item \emph{Gain smoothing}: we apply a smoothing filter to the recovered gain solutions across frequency to reject spurious features caused by the inherent nonredundancies in the data.
    \item \emph{Temporal visibility filtering}: we apply the temporal filter technique \citep{Parsons2009, Charles2023}, or fringe-rate filtering, which filters the data prior to calibration to down-weight strong nonredundant features in the data originating from galactic diffuse emission entering the side lobes of the primary beam.
\end{itemize}
Note that Charles et al. (2023 in prep) also discuss the use of temporal filtering to suppress nonredundant errors in calibration, but they focus on its application in mitigating mutual coupling systematics, whereas this work focuses solely on nonredundancies caused by per-antenna feed displacements.

In Section~\ref{sec:vis_sim}, we explain about the visibility simulation and the input gain model. Section~\ref{sec:redcal} describes the redundant baseline calibration scheme, followed by the absolute calibration, and presents the calibration gain errors before mitigation. Section~\ref{sec:mitigation} discusses the effects of three different mitigation methods and their combinations on reducing the chromatic gain errors. The power spectrum analysis without and with mitigation is performed in Section~\ref{sec:pspec}. Section~\ref{sec:conclusions} summarizes the main results of this study. Throughout the paper, we adopted the cosmological parameters from \citet{Planck2016} which are consistent with what we used in Paper~I: $\Omega_\Lambda = 0.6844$, $\Omega_{\rm m} = 0.3156$, $\Omega_{\rm b} = 0.04911$, and $H_0 = 67.27\,{\rm km\, s^{-1}\, Mpc^{-1}}$.

\section{Visibility Simulations with Perturbed Primary Beams}
\label{sec:vis_sim}
As described in Paper I, we simplified the feed motions by separating them into vertical, horizontal, and tilting offsets. We utilized a three-dimensional Vivaldi feed model for the calculation of the primary beam at each feed position using the CST electromagnetic simulation software. The fiducial model locates the feed at the place, 5~m above the vertex of the 14-m parabolic dish \citep{Fagnoni2021}. For the perturbed feed models, we displaced the feed away from the fiducial position and simulated far-field electric fields. The far-field electric fields were calculated for the 160--180~MHz band, sampled at 0.125~MHz. The behavior of the primary beam responses alters for different feed perturbations. The vertical feed offset mainly causes a change to the width of the main lobe of the primary beam, while the horizontal or tilting ones mainly results in a change to the pointing angle of the main lobe.

We utilized the HERA core array configuration consisting of 320 antenna elements for visibility simulations \citep[HERA-320;][]{Dillon2016}. The HERA array is designed to be compact with a large number of redundant baselines, so that it can achieve high-precision calibration and high sensitivity to the 21\,cm power spectrum. The array configuration requires 51,360 visibility calculations per frequency bin and per time integration.

The visibility measurement for a given antenna pair, frequency, and time is calculated as follows\footnote{We developed a visibility simulator for fast computation of visibilities using multiple beam models that are randomly perturbed. It is publicly available in \url{https://github.com/vispb/vispb}.} \citep[e.g.,][]{Smirnov2011}.
\begin{equation}
V_{ij}(\nu, t) = \sum^{N_{\rm src}}_{n=1} B_{ij}(\hat{s}_n(t), \nu) S(\hat{s}_n(t), \nu) e^{-2\pi i\nu \mathbf{b}_{ij} \cdot \hat{s}_n(t)/c},
\label{eqn:vis_sim}
\end{equation}
\noindent where $V_{ij}$ is the complex visibility at time $t$ and frequency bin $\nu$, $\textbf{b}_{ij}$ is a baseline vector between antennas $i$ and $j$, and $\hat{s}_n(t)$ is a pointing vector to a source varying with time in the local observer frame. We consider the middle band of HERA, 160--180~MHz, corresponding to the redshift of $\sim$7 at a cadence of 0.25~MHz. There are three components involved with the calculation. $B_{ij}$ is the power beam computed from far-field electric fields of the antenna pair $i$ and $j$. $S$ is the flux density of a sky source in Jy that corresponds to either a compact source or a discretized pixel of a diffuse source. The last exponential term is the baseline interferometric fringe response.

The power beam term is defined by $B_{ij}(\theta, \phi, \nu) = E^p_{i,\theta} {E^p_{j,\theta}}^* + E^p_{i,\phi} {E^p_{j,\phi}}^*$ where $p$ is the feed polarization \citep[e.g.,][]{Kohn2019}. We use an East-oriented feed polarization throughout the study. Each far-field electric field of antenna $i$ and $j$ is perturbed by the feed motions. We consider vertical, horizontal, and tilting motions separately. Those motions follow a Gaussian distribution with zero mean and standard deviation $\sigma_{\rm feed}$~=~2, 3, and 4~cm for the vertical and horizontal feed motions and $\sigma_{\rm feed}$~=~2$^\circ$ and 3$^\circ$ feed tilting at the fiducial feed position, providing 8 different simulation sets. The perturbed power beam responses were then fed into the sky visibility calculation. Descriptions of the beam simulations and visibility simulations are detailed in Paper~I.

For the foreground emission, we consider two different sky components, a point source model and a diffuse sky model. The former comes from the GaLactic and Extragalactic All-sky MWA (GLEAM) survey. We combined GLEAM~I \citep{Hurley-Walker2017} and GLEAM~II \citep{Hurley-Walker2019} point source catalogs as well as peeled bright sources listed in Table~2 of \citet{Hurley-Walker2017} and Fornax A \citep[e.g.,][]{Bernardi2013}. The diffuse sky model is constructed with Hierarchical Equal Area isoLatitude Pixelization \citep[\texttt{HEALPix};][]{Gorski2005} using the Global Sky Model \citep[GSM;][]{Zheng2017} that removed the bright compact sources from the sky model. This makes us minimize double counting bright point sources when we combine the point source model and the diffuse sky model. We chose \texttt{HEALPix} \texttt{Nside}~=~256, containing 786,432 pixels with the pixel area of $1.60\times10^{-5}$~sr. When the pixel area is larger than the angular resolution of a particular baseline ($\theta\sim\lambda/|\mathbf{b}|$), an error may be introduced into the power spectrum due to under-sampling of the discretized map \citep{Lanman2020}. For the array configuration and the frequency band we adopted, the angular resolution corresponding to the longest baseline, 292~m, is $\sim3\times10^{-5}$~sr that satisfies the Nyquist limit. This means for most of baselines, especially for short baselines that could achieve high sensitivity in power spectrum estimate, the adopted pixel area of the GSM does not have a critical impact on our power spectrum analysis.

The cosmological signals are simulated as a reference for the detection of the EoR. To simulate the EoR signal, we employed the \texttt{21cmFAST} public software \citep{Mesinger2011} with the fiducial model studied by \citet{Park2019}. The cube size and the number of pixels per side are 250~Mpc and 128, respectively. The cube is simulated at the redshifts of interest spanning 160--180~MHz, and each cube is converted to the \texttt{HEALPix} map to feed into the visibility simulation by following the method described in \citet{Kittiwisit2017}. For the conversion, we considered \texttt{Nside} = 8192 so that at least two \texttt{HEALPix} pixels are sampled at each cube cell, and degraded the resolution of the \texttt{HEALPix} to \texttt{Nside} = 256 that is computationally feasible for the visibility simulation. The EoR visibility is simulated with the fiducial primary beam. Because the EoR visibility is only for assessing the detectability of the EoR in the power spectrum analysis, no gains and calibration are applied.

As the sky rotates with time, the nonredundancy effect can alter at a different local sidereal time (LST). For example, Paper I found that when strong diffuse emission, such as the galactic plane, is located near the horizon, the deformation of side lobes due to feed motions can induce significant nonredundancies in the visibility data. Additionally, another type of nonredundancy can arise when strong compact radio sources come in and out the main lobe with time. This effect is expected to be stronger for the vertical feed motion since it predominantly affects the beam width of the main lobe comparing to the horizontal and tilting perturbations.

In this work, we simulated the visibility over LST of 0--3 hours at a cadence of 50~seconds. The range and sampling rate along the LST is chosen for the temporal filter defined in the fringe rate ($f$), carrying units of millihertz (mHz), the Fourier dual of time defined as $\tilde{V}(f) = \int dt\ e^{2\pi i tf}V(t)$. That is, we have enough resolution ($\delta$f $\sim$ 0.046~mHz) and coverage ($\Delta$f $\sim$ 10~mHz) in the fringe-rate domain to perform the fringe-rate filter which will filter out signals $< 0.3$~mHz. Throughout the analysis, the feed positions are fixed, and we do not take into account time-dependent feed motions.

The simulated raw visibilities are corrupted by multiplying complex antenna gains to yield the ``observed" visibilities, $V_{ij}^{\rm obs} = g_i g_j^*V_{ij}^{\rm raw}$. The antenna gains $g_i$ are modeled by following the calculation of the voltage response described in Equation of (5) of \citet{Fagnoni2021}. The calculation is based on the impedance of the antenna, the electric-field pattern, and the scattering parameter. The gain model is interpolated by using Gaussian process regression (GPR) with a ﬁxed kernel size of 1~MHz, which effectively smooths out frequency structure smaller than 1~MHz \citep[e.g.][]{Kern2018, Lanman2020}. Then the input gains are randomly perturbed in amplitudes ($\eta_i)$ and cable delays ($\tau_i$), by exp$(\eta_i + 2\pi i \tau_i \nu)$, according to a Gaussian distribution of zero mean and a standard deviation of 0.2 and 20~ns, respectively. Details of the full antenna gain derivation are described in Paper I. We ignore thermal noise in generating the measured visibilities to focus on the foreground contamination leaking into the EoR window that are supposed to be foreground-free.

\section{Redundant-baseline Calibration and Observed Chromatic Gain Errors}
\label{sec:redcal}

\begin{figure*}[t!]
\centering
\includegraphics[scale=0.5]{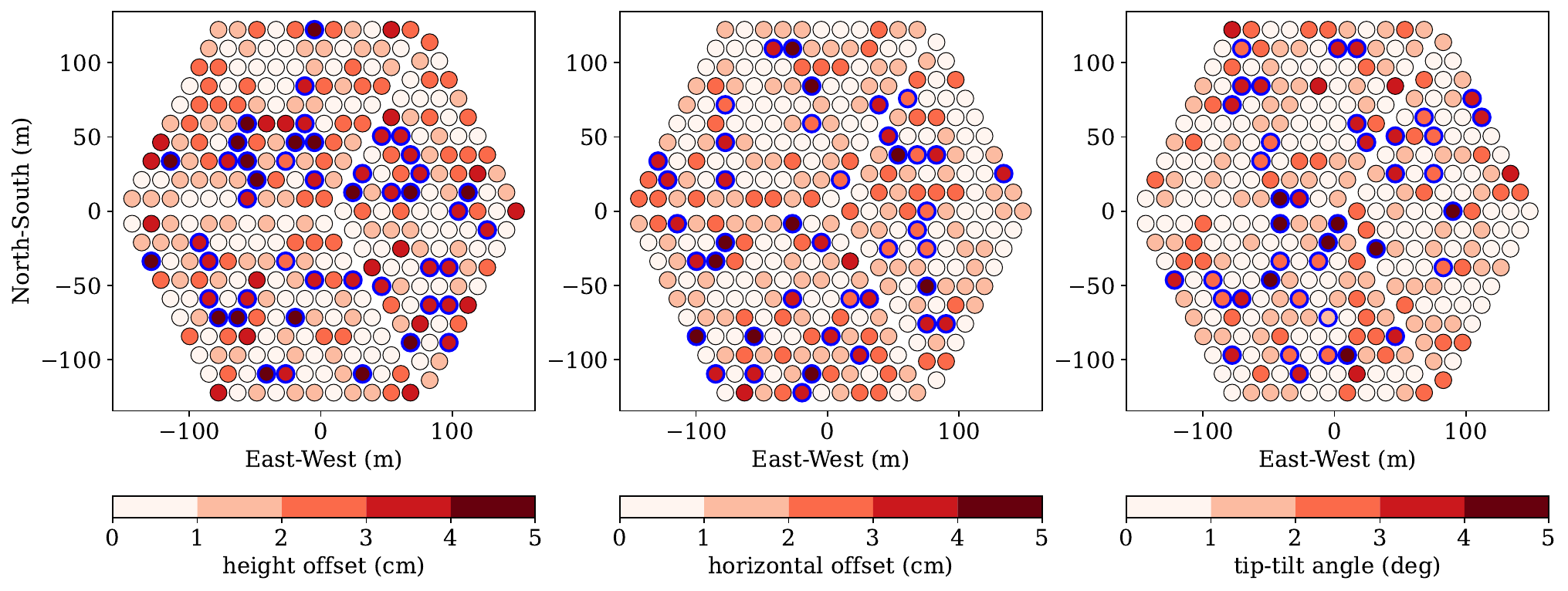}
\caption{Antenna layout with antennas colored according to the feed offset. This is an example when the feed positions are perturbed by following a Gaussian distribution with zero mean and $\sigma_{\rm feed} = 2$~cm for the vertical and horizontal feed motions and $\sigma_{\rm feed} = 2^\circ$ for the tilting feed motion. The color bar indicates absolute values of the offsets that are sampled to 5 bins for illustrative purposes. The flagged antennas based on the modified Z-score metric in the redundant-baseline calibration are marked with blue circular boundaries. Overall, the metric works well enough to exclude bad antennas with large offset.}
\label{fig:layout_bad_ants}
\end{figure*}

HERA benefits from the highly redundant baseline configuration for precise calibration. We follow the standard HERA direction-independent calibration scheme consisting of a redundant baseline calibration step, followed by an absolute calibration step that solves for the inherent degeneracies in redundant calibration. Instead of using prior information of the sky, the redundant baseline calibration uses the redundancy of measurements of the same baselines. For antenna pairs with the same baseline separation, they are assumed to share the same unique visibility solution, $V_{i-j}^{\rm sol}$. Namely, the redundant calibration is carried out by solving $V_{ij}^{\rm obs} = g_i g_j^*V_{i-j}^{\rm sol}$ to obtain $g_i$, $g_j$, and $V_{i-j}^{\rm sol}$ by minimizing the $\chi^2$ statistic,
\begin{equation}
    \chi^2 = \sum_{i<j} \frac{|V_{ij}^{\rm obs} - g_i g_j^* V_{i-j}^{\rm sol}|^2}{\sigma_{ij}^2},
    \label{eqn:chi2}
\end{equation}
where $\sigma_{ij}^2$ is the thermal noise corresponding to each $V_{ij}^{\rm obs}$.

The antenna gains and unique visibility solutions derived from the redundant baseline calibration can be degenerate. There are four well-known degenerate parameters, including the overall amplitude ($g_i \rightarrow \hat{A} g_i$, $V_{i-j}^{\rm sol} \rightarrow V_{i-j}^{\rm sol}/\hat{A}^2$), the overall phase ($g_i \rightarrow e^{i\psi} g_i$, $g_j^* \rightarrow e^{-i\psi} g_j^*$), and two phase gradients associated with tip-tilt slopes across the array along the east-west ($g_i \rightarrow g_i e^{i\Phi_x x_i}$ and $V_{i-j}^{\rm sol} \rightarrow V_{i-j}^{\rm sol}e^{-i\Phi_x (x_i-x_j)}$) and the north-south direction ($g_i \rightarrow g_i e^{i\Phi_y y_i}$ and $V_{i-j}^{\rm sol} \rightarrow V_{i-j}^{\rm sol}e^{-i\Phi_y (y_i-y_j)}$) where $\hat{A}$, $\psi$, $\Phi_x$, and $\Phi_y$ are scalar free parameters and $x_i$ and $y_i$ are antenna positions \citep{Liu2010, Zheng2014, Dillon2018, Li2018, Byrne2019, Dillon2020, Kern2020a}. To solve for these degeneracies, an absolute calibration step using sky information is needed. In this second calibration step, the raw visibility simulated with the unperturbed beam is chosen for the model visibility. As discussed in Paper I, when the feed positions are perturbed, majority of the chromatic gain errors come from the redundant baseline calibration step, and the absolute calibration mainly causes the errors at high delays.

During the calibration, we exclude bad antennas with relatively larger feed displacement that may degrade calibration solutions. The process is implemented by ruling out antennas with a modified Z-score $> 4$ in the redundant-baseline calibration. The modified Z-score for antenna $i$, described in \citet{Kern2022}, is defined as $Z_i^{\rm mod} = (x_i - {\rm med}(x))/\sigma^{\rm mad}$ where $\sigma^{\rm mad} = 1.482 \times {\rm med}|x-{\rm med}(x)|$ which is the median absolute deviation. $x_i$ is the mean visibility amplitude computed by $x_i=\sum_{j \ne i, \nu, t}|V_{ij}(\nu, t)|/(N-1)$ where $N$ is the number of antennas and the sum is performed over $j \ne i$, $\nu$, and $t$. This metric takes an iterative process that is repeated three times with the removal of bad antennas (i.e., modified Z-score $> 4$) in every iteration.

\begin{figure*}[t!]
\centering
\includegraphics[scale=0.55]{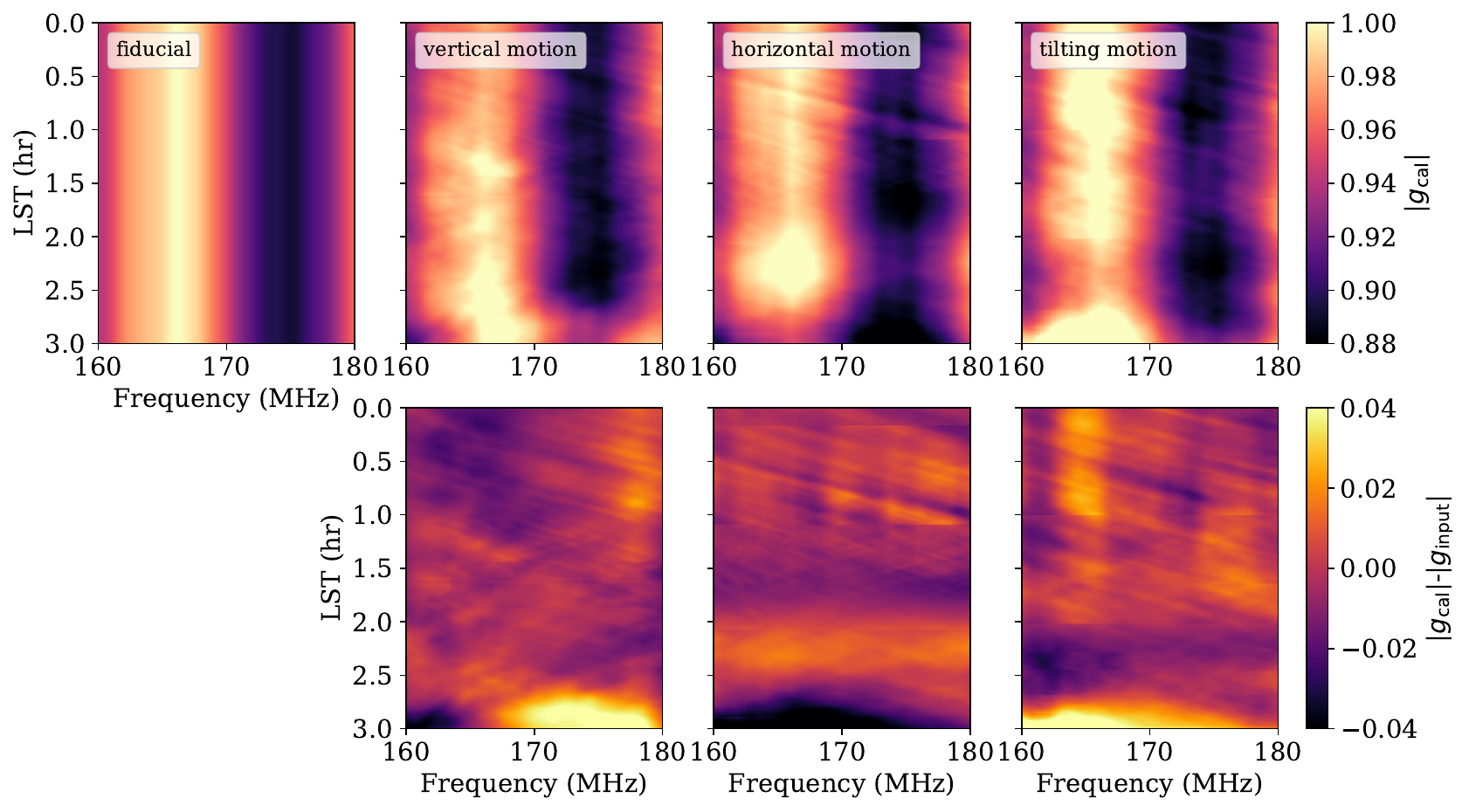}
\caption{Antenna gain solutions derived from the perturbed beam simulations with LST compared to the fiducial one. Three randomly perturbed antennas are picked from separate sets of the simulations for vertical, horizontal, and tilting motions, respectively. Each antenna is perturbed with $-4.3$~cm of height offset (center left), $+3.4$~cm of east-direction displacement (center right), and $+3.1^\circ$ of east-direction tip-tilt (far right). Top panels show the gain amplitudes and the bottom panels show the difference between the calibrated gains and the input gains. For the fiducial model, the gain solutions have constant temporal structure as the input gains without any artificial features. When the feed is perturbed, fine-scale artifacts show up and result in corrupting the cosmological modes in the EoR window.}
\label{fig:gain_errors_with_LST}
\end{figure*}

Figure~\ref{fig:layout_bad_ants} illustrates how the modified Z-score works to exclude bad antennas during the calibration process. The feed is displaced according to a Gaussian distribution of the zero mean and $\sigma_{\rm feed} = 2$~cm for the translation feed motion and $\sigma_{\rm feed} = 2^\circ$ for the tilting feed motion. The flagged antennas with the modified Z-score $> 4$ are highlighted with bold edges of blue circles. This result indicates that many antennas with large feed offset around 3--5~cm or 3--5$^\circ$ are ruled out during the calibration and the modified Z-score metric is an effective tool to identify bad antennas that may harm the calibration. We found that this reduces the overall $\chi^2$ defined in Equation~\eqref{eqn:chi2} but has little impact on reducing spurious spectral gain structure and thus no advance in terms of reducing the foreground power leakage in the power spectrum.

In practice, antennas that contribute large errors to the calibration solution are flagged. To be consistent with actual HERA calibration techniques, calibration solutions in this analysis are calculated with flagged antenna omitted. We only consider $\sigma_{\rm feed} = 2$~cm for the axial and lateral feed motions and $\sigma_{\rm feed} = 2^\circ$ for the tip-tilt motion throughout the paper unless it is explicitly stated otherwise.

\subsection{Chromatic Gain Errors}

For the fiducial model where all antenna elements have identical beam responses, the calibration is expected to be perfect without introducing additional frequency structure to the antenna gain solutions. When the feeds are perturbed, however, the key assumption of redundancy is violated, and the mismatch between visibilities in the redundant baseline group may introduce artifacts to the gain solution to account for the nonredundancy in the data. In Figure~\ref{fig:gain_errors_with_LST}, we demonstrate how the artifacts look like for the vertical (second column), horizontal (third column), and tilting (forth column) feed motions compared to the fiducial one (first column). We select three antennas to illustrate the effects of different types of feed motions on the recovered gain solutions: one antenna displaced by $-4.3$~cm in height from the vertical feed motion simulation, another off by $+3.4$~cm in east-direction from the horizontal motion simulation, and the other perturbed by $+3.1^\circ$ in east-direction from the tip-tilt simulation.

\begin{figure*}[t!]
\centering
\includegraphics[scale=0.55]{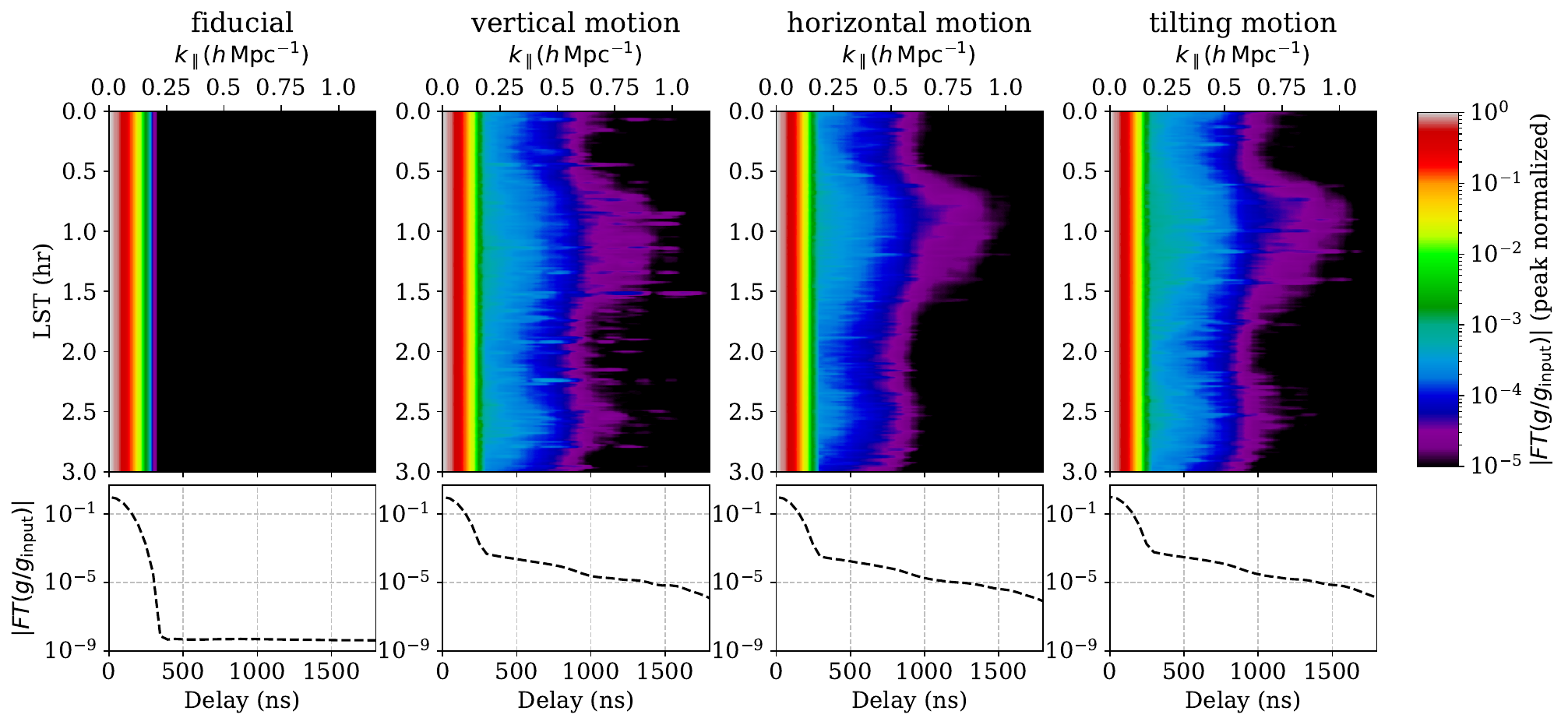}
\caption{Delay spectrum of the calibrated gains without mitigation. Top panels show the gain delay spectrum varying with LST. Bottom panels present the LST-averaged delay spectrum. The fiducial beam model (far left panel) recovers the input gain, providing the clean window at high delay modes $> 400$~ns. $g/g_{\rm input}$ is constant with frequency and the bell-shaped curve shown in the bottom panel comes from the seven-term Blackman-Harris tapering function. For the perturbed cases, the window $\kpara > 0.25 \, \hMpcinv$ is corrupted by chromatic gain errors arising from the nonredundancy in measurements. The bulge occurring at 0.5--1.5~hr is due to the galactic plane located near the horizon.}
\label{fig:fft_gain_errors_no_mitigation_LST}
\end{figure*}

As the input gain does not change in time, we expect calibration gains from the fiducial beam model should also be time-constant as shown in the first panel of Figure~\ref{fig:gain_errors_with_LST}. When feeds are perturbed, regardless of the type of feed motions, calibration solutions show additional spectral and temporal structure. The artificial features are evident in the difference between the calibration gains and the input gains (bottom row).

There are two different scales of artifacts. One is the change in the bulk shape of the gain that can be understood based on the primary beam patterns changing with the feed motions. For example, in the bottom panel of the vertical motion, there are negative and positive features along frequency at LST$\sim$3~hr that is close to Fornax~A located at 3.3~hr, one of the brightest radio sources in the HERA stripe \citep[e.g.,][]{Dillon2020}. Paper~I shows that the change in the feed height can drive the change to the beam width. With the height offset of $-4.3$~cm, the beam width is narrower than the fiducial beam at 160~MHz and gets broader as the frequency increases, and reaches a similar beam width level as the fiducial one at $\sim$165~MHz (see Figure~3 of Paper~I). Such a behavior indicates the flux density of Fornax~A is more/less attenuated by the primary beam than the fiducial one at frequency below/above 165~MHz, which is consistent with the pattern shown in Figure~\ref{fig:gain_errors_with_LST}. 

Similarly, when the feed moves in the horizontal direction or tilts, the primary beam's center position shifts. According to the results of Paper~I, when the feed moves in the east-direction, the beam shifts in the opposite direction, and hence the beam-weighted flux density of Fornax~A becomes dimmer than the fiducial case at LST$\sim$3~hr, resulting in smaller gain amplitudes. Meanwhile, when the feed tilts in the east-direction, the beam shifts in the same direction, and the beam-weighted flux density of Fornax~A is brighter than the fiducial one at LST$\sim$3~hr, leading to larger gain amplitudes than the fiducial beam model. Additionally, there are also horizontal patterns between 1.5 and 2.5~hr in the bottom panels of horizontal and tilting motions. Those are caused by two point sources transiting zenith at LST$\sim$2~hr, and can be understood in the same way as the LST$\sim$3~hr feature.

Another noticeable artificial features are fine-scale structure in frequency. Because we include long baselines in the calibration, artificial gain features arising from the baselines are expected to form fine structure in frequency, resulting in complex patterns as observed in Figure~\ref{fig:gain_errors_with_LST}. These chromatic gain errors corrupting high delay modes are ones responsible for the foreground leakage observed in Paper~I. Figure~\ref{fig:fft_gain_errors_no_mitigation_LST} presents how the chromatic gain errors of the perturbed beam models are shaped in the delay domain compared to the fiducial model. To transform the gain to the delay spectrum, we divided the gain by the input gain, multiplied it by a seven-term Blackman-Harris tapering function along the frequency axis, and performed the frequency Fourier transform. Then we averaged the amplitudes of the peak-normalized delay spectra over unflagged antennas at each LST bin. In the first column, we show results of the fiducial model with the time-constant gain solutions (top) and the LST-averaged gain solution consistent with the input gain as expected (bottom).

\begin{figure}[t!]
\centering
\includegraphics[scale=0.55]{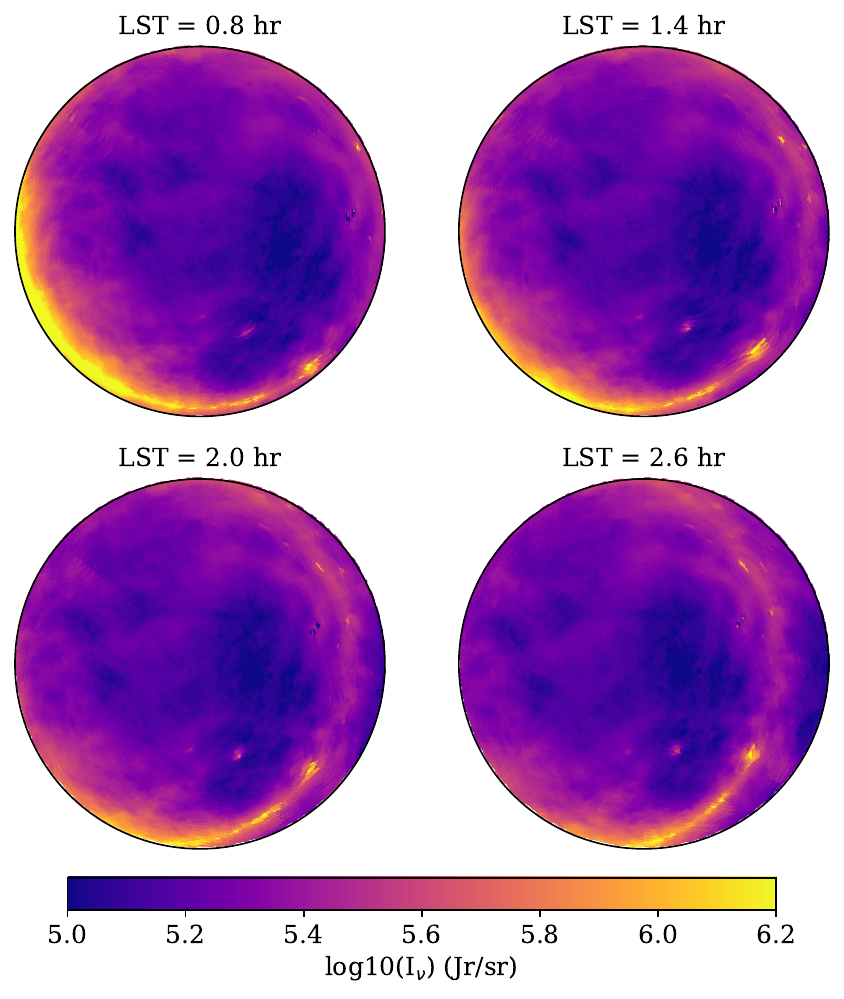}
\caption{GSM maps viewed in horizon coordinates from the HERA observing site at ($-30.72^\circ$S, $21.43^\circ$E), plotted at different local sidereal times. At LST$\sim$0.8~hr, the galactic plane lies near the horizon, and the bright emission disappears as the Earth rotates.}
\label{fig:GSM_with_LST}
\end{figure}

As shown in Figure~\ref{fig:fft_gain_errors_no_mitigation_LST} from the second left to right columns, when the feeds are displaced from the fiducial position, gain errors bleed into high delay modes. The overall patterns are similar regardless of the type of feed motions, showing more power leakage than the fiducial model across all LST. The strongest bleeding regions appear to be at LST range of 0.5--1.5~hr, when the galactic plane lies near the horizon and sets as the Earth rotates as depicted in Figure~\ref{fig:GSM_with_LST}. Since the perturbed primary beams induced by the feed motions are responsive to the strong diffuse emission from horizon, it can be understood that the bulge produced at LST$\sim$1 hr is most likely due to the galactic plane near the horizon. In addition, because the feed offsets, especially vertical ones, are sensitive to the stochastic distribution of point sources, they introduce irregularly narrow structures across each LST bin.

Bottom panels of Figure~\ref{fig:fft_gain_errors_no_mitigation_LST} show the averaged gain delay spectra over LST. For the perturbed feed models, power in the delay spectrum is 4--5 orders of magnitudes stronger than the input gain at 500--1000~ns. Because the power spectrum of cosmological signals is expected to be $10^{10}$ times smaller than the foregrounds \citep[e.g.,][]{Thyagarajan2016}, the delay spectrum should be smaller than $10^{-5}$ at around 700~ns or $\kpara\sim0.4$~$\hMpcinv$, which indicates we need to mitigate the chromatic gain errors to achieve robust detection of the EoR\footnote{Though averaging uncorrelated gain errors in forming a power spectrum may lessen this stringent requirement of $10^{-5}$ in gain powers, we use the requirement as it guarantees the recovery of a clean EoR window as shown in Figure~\ref{fig:2d_pspec_mitigated}.}. We chose $\kpara\sim0.4$~$\hMpcinv$ since it provides a foreground-free region for the fiducial power spectrum (see Figure~\ref{fig:pspec_fiducial}) with high sensitivity when thermal noise is included. In the following section, we introduce three different ways of mitigating these chromatic gain errors.

\section{Mitigation of the spectral structure in the gain solutions}
\label{sec:mitigation}

In this section, we explore three different ways of mitigating spurious fine-scale frequency structure in the derived antenna gains due to nonredundancy created by random feed displacements. Two of the methods, baseline cut and temporal filtering, treat the visibilities before calibration, thus mitigating potential errors introduced by the calibration process itself. Meanwhile, the other method, gain smoothing, treats the derived gains after performing calibration. Baseline cut is to down-weight long baselines in the data, which contribute to gain errors at fine spectral scales during calibration. Gain smoothing is to suppress the fine-scale frequency structure in the derived gain solutions using a low-pass frequency filter after the calibration, thus smoothing the gains. Temporal filtering is to apply the fringe-rate notch filter that excludes the horizon emission before performing the redundant calibration. This is effective to reduce the nonredundancy in the visibility when strong emission such as the galactic plane is lying on the horizon. All different strategies are discussed separately and then combined to illustrate the composite effect of the mitigation on the power spectrum in Section~\ref{sec:pspec}.

\begin{figure*}[t!]
\centering
\includegraphics[scale=0.5]{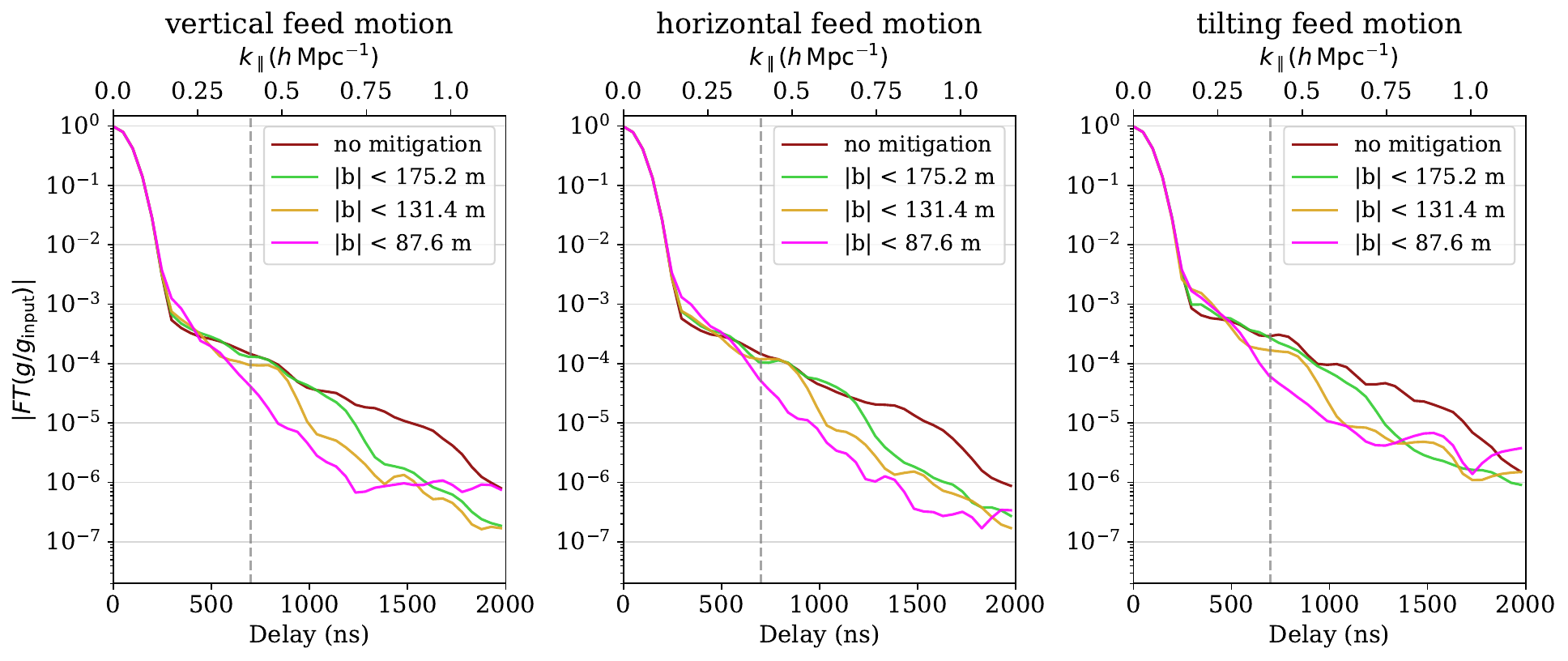}
\caption{The gain delay spectra with different maximum baseline cut-offs for vertical (left), horizontal (middle), and tilting feed motions (right) at 1~hr of LST. The Fourier transform of the antenna gain divided by the true input gain is averaged over the unflagged antennas. There is a trend that the broad wing appearing at high delays gets reduced as we consider tighter baseline limits for the calibration, indicating the chromatic gain errors are mitigated with the restriction of the baseline lengths. There is an order of magnitude improvement at 700~ns or $k=0.4\,\hMpcinv$ (vertical dashed line) for $|\mathbf{b}| < 87.6$~m case compared to no mitigation case.}
\label{fig:fft_gain_mitigation_bl_cut}
\end{figure*}

\begin{figure*}[t!]
\centering
\includegraphics[scale=0.55]{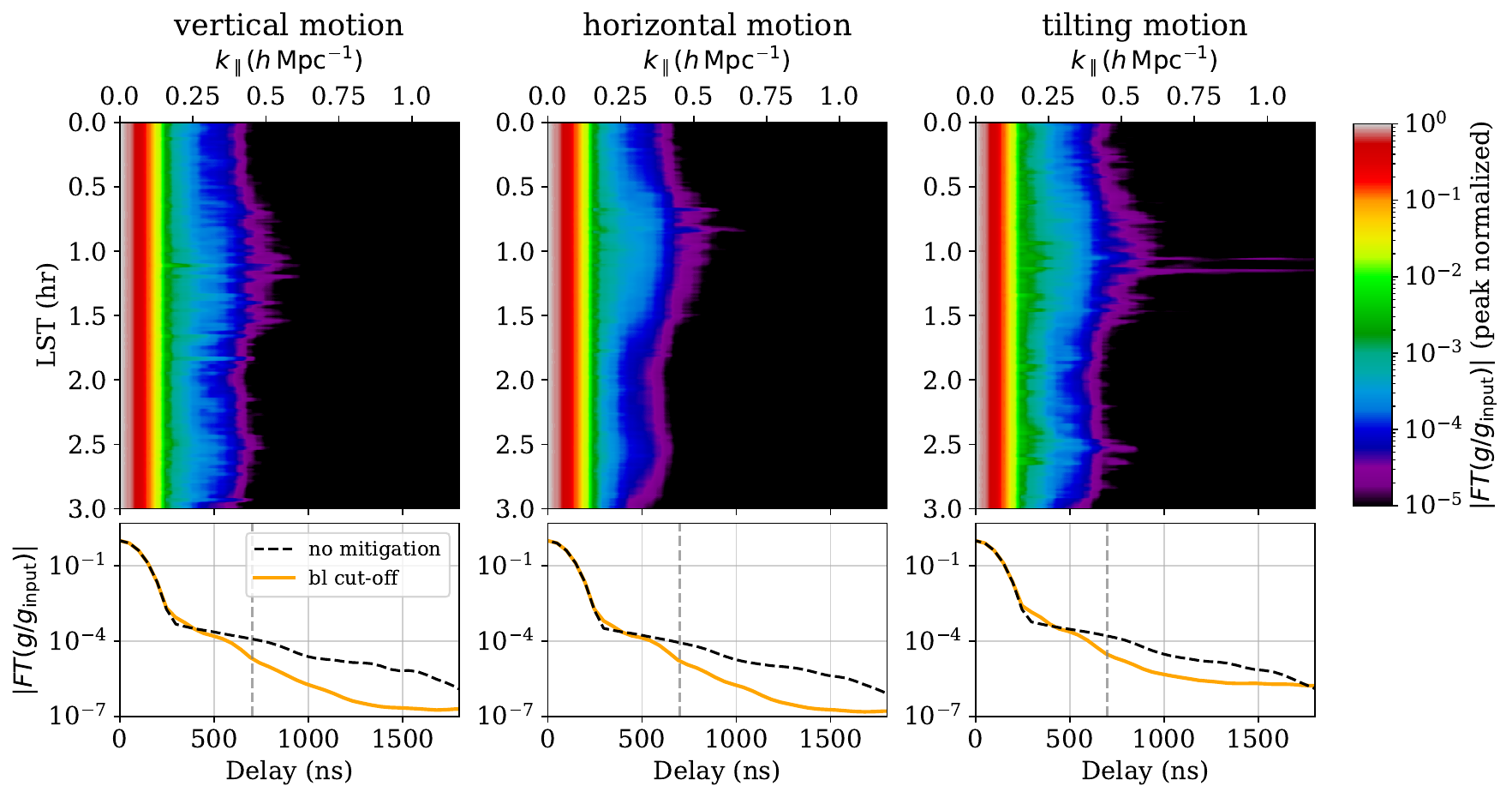}
\caption{Delay spectrum of the gain solutions with down-weighting long baselines. We consider 87.6~m baseline cut-off for the mitigation. The top row shows the delay spectrum for the mitigated gain with LST for the vertical (left), horizontal (middle), and tilting (right) feed motions. Overall, the delay spectra fall down to $< 5\times10^{-5}$ at 700~ns, which is also manifested in the bottom row for the LST-averaged delay spectra.}
\label{fig:fft_gain_errors_bl_cut_LST}
\end{figure*}

\subsection{Down-weighting Long Baselines}
\label{sec:baseline_cutoff}
High-frequency errors in antenna gains caused by nonredundancy associated with compact radio sources can predominantly arise from long baselines with high delays ($|\mathbf{b}|/c$). For example, visibilities suffering from nonredundancy in a 200-m baseline group, which has a delay of $\sim$700~ns, can be affected by nonredundancy errors at the scale of $\sim$1~MHz or smaller, which will propagate into the same scale of gain errors during the calibration process.  \citet{Ewall-Wice2016} and \citet{Orosz2019} found that chromatic gain errors can be reduced by placing a restriction on baseline lengths used in the redundant calibration. Following the approach of \citet{Orosz2019}, we considered a binary weight to down-weight baselines that are longer than certain threshold (i.e., 1 for shorter baselines and 0 otherwise).

In order to look for an optimal maximum baseline cut-off, we examined the improvement of calibration gains by changing the baseline cut-off used for the redundant calibration and absolute calibration. For the baseline cut-off values, we chose 87.6, 131.4, and 175.2~m which correspond to 6$\times$, 9$\times$, and 12$\times$ the shortest baseline (14.6~m), respectively.\footnote{In practice, we also impose a minimum baseline cut-off to reduce the effect of modeling uncertainty of diffuse emission in the absolute calibration. Thus, choosing maximum baseline cut-off smaller than 87.6~m  may cause lack of redundant baselines in the system.} Figure~\ref{fig:fft_gain_mitigation_bl_cut} presents the gain delay spectra with different baseline cut-offs for the vertical (left), horizontal (middle), and tilting feed motion (right) at LST~=~1~hr. Each curve is computed as the Fourier transform of the ratio of antenna gains and input gains, weighted by a seven-term Blackman-Harris window function, then averaged over all unflagged antennas. Regardless of the types of feed motions, it is apparent that the power leakage at high delays is suppressed more as the baseline limit becomes tighter, which supports the finding of \citet{Orosz2019}.

\begin{figure*}[t!]
\centering
\includegraphics[scale=0.55]{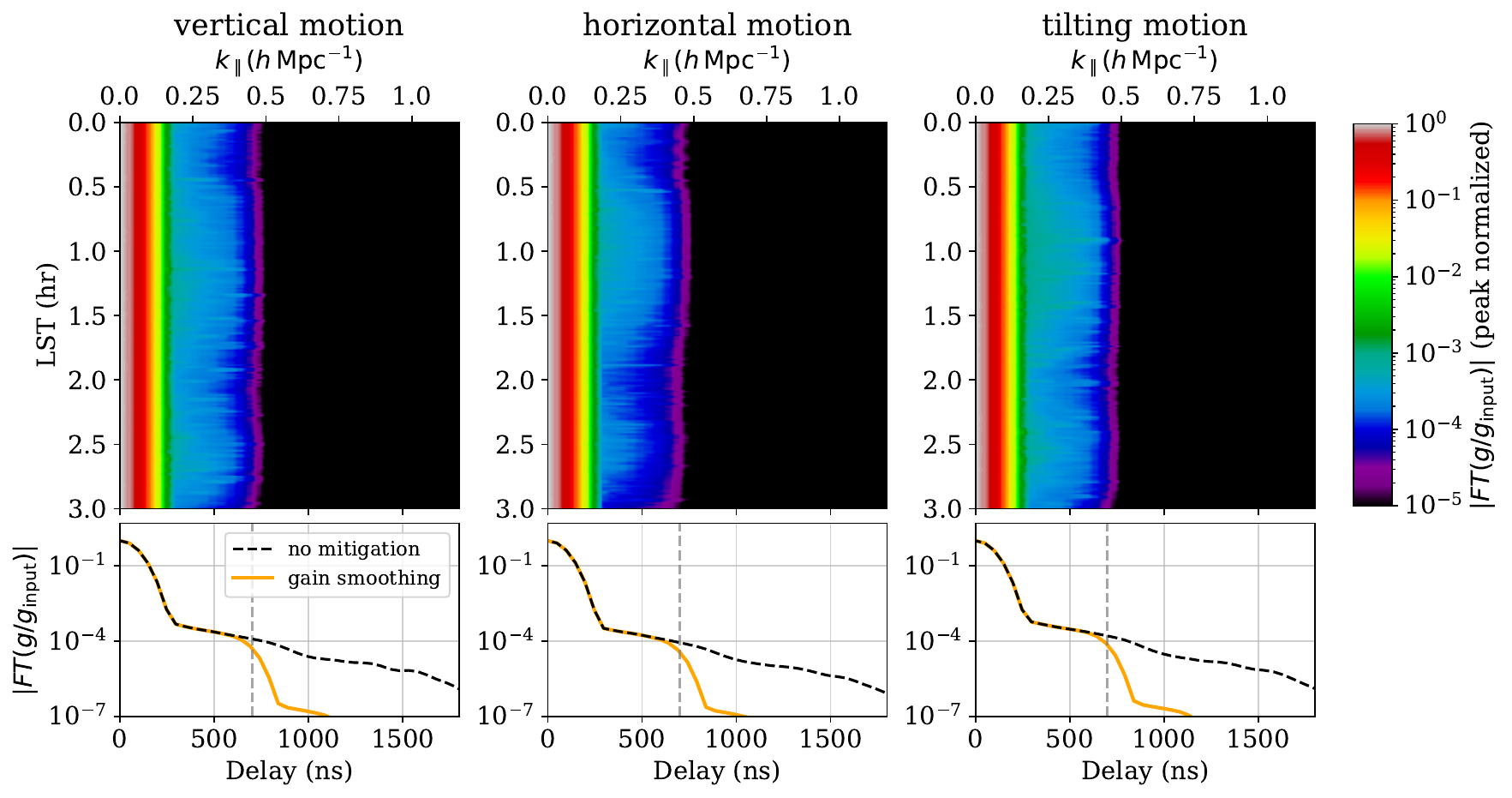}
\caption{Delay spectrum of the gain solutions with gain smoothing along the frequency. The format is the same as that of Figure~\ref{fig:fft_gain_errors_bl_cut_LST}. As the Fourier filter rules out the spectral scales larger than 1~MHz, or smaller than 500~ns, the top panels show the clean window beyond 700~ns for all types of the feed motions over all LSTs as expected.}
\label{fig:fft_gain_errors_frequency_smoothcal_LST}
\end{figure*}

With the cut-off of 87.6~m, the power leakage is reduced by an order of magnitude at $\kpara\sim0.4$~$\hMpcinv$ (vertical dashed line). For all cases, the mitigation brings the power leakage down to $5\times10^{-5}$, which is still a bit higher than desired. There is an additional bump at $\sim$300~ns (magenta curve) compared to no mitigation case since redundant calibration is most sensitive to the nonredundancy errors arsing from the cut-off baseline corresponding to the delay of 300~ns (i.e., $87.6/c \sim$ 300~ns).

In Figure~\ref{fig:fft_gain_errors_bl_cut_LST}, we illustrate how the mitigation with the 87.6-m cut-off is effective over the LST range. Across all LSTs, the mitigated gain has reduced spectral artifacts compared to the case without mitigation. The mitigation is relatively less effective at LST between 0.5 and 1.5~hr, where significant chromatic gain errors result from strong horizon emission captured by the perturbed beams. This is even worse for the tilting motion as there are more variations in the side lobes owing to that motion than other motions. The LST-averaged delay spectra in the bottom row of Figure~\ref{fig:fft_gain_errors_bl_cut_LST} indicate that the chromatic gain errors are sufficiently mitigated at high delays but need to be suppressed further at $\kpara = 0.4 \,\hMpcinv$. We adopt 87.6~m as the optimal baseline cut-off for the remaining sections.

\subsection{Gain Smoothing}

The mitigation of fine-scale spurious features in the derived antenna gain solutions can also be achieved by smoothing the gains in some manner. \citet{Barry2016} proposed doing this with low-order polynomials, whereas \citet{Kern2020a, Kern2022} propose doing this with Fourier filters, while \citet{Yatawatta2015} enforce spectral smoothness in the process of calibration itself.
In this work, we use a formalism similar to the one used in \citet{Kern2022} for smoothing the derived gains in frequency with a low-pass filter. Our input gain is designed to have features at a scale larger than 1~MHz. Thus, we find that a 1-MHz smoothing scale to be the most effective in rejecting spectral gain errors while retaining major components of the true gain structure that need to be calibrated out.

The delay spectrum results after applying the low-pass filter with the filter size of 500~ns, corresponding to the 1~MHz smoothing scale, are shown in Figure~\ref{fig:fft_gain_errors_frequency_smoothcal_LST}. As expected, power leakage beyond 700~ns (vertical dashed line) is well suppressed below the level of $10^{-5}$ across all LST bins for all three types of feed motions. However, unlike other mitigation methods used in this study, this method is essentially a low-pass filtering thus, by design, ineffective in reducing gain error power leakage at delays $<500$~ns, which corrupts low $k$-modes in the power spectrum. The LST-averaged delay spectra for all feed offset types (bottom panels) also display a fast drop-off of the power beyond 700~ns, and no mitigation of the chromatic errors inside 500~ns.

\begin{figure*}[t!]
\centering
\includegraphics[scale=0.5]{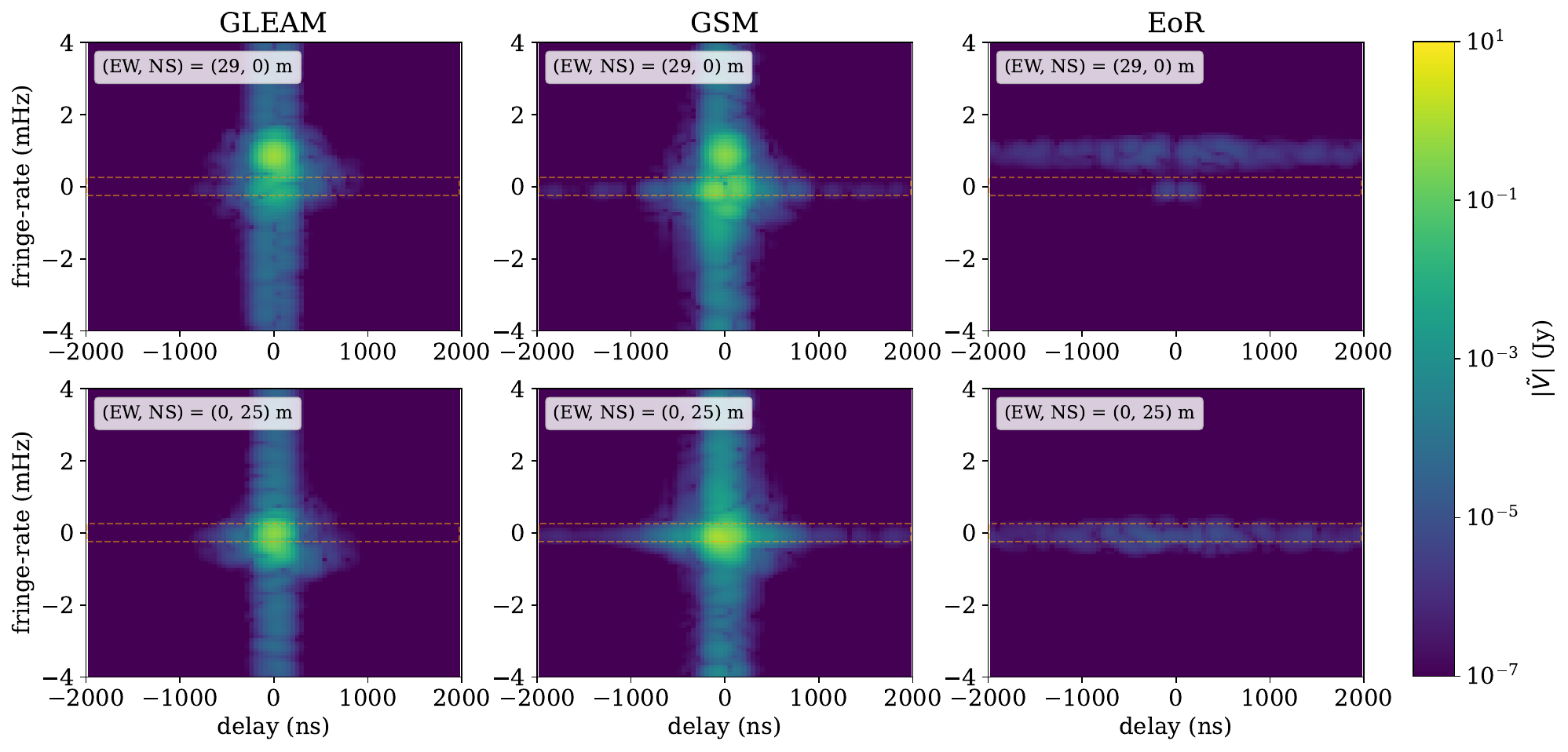}
\caption{Fourier transform of simulated HERA visibilities for GLEAM (left), GSM (middle), and EoR (right) sky models in the delay and fringe-rate domain. Two antenna pairs with similar separations but with different orientations are presented as labeled in the plot. For the north-south only baseline (bottom row), all types of sky models have their power located near zero fringe rates, which drift away from the center as the baseline length in the east-west direction increases. On the top row, the two central bloblike features in the GSM panel that originate from the pitchfork effect of the horizon emission. The fringe-rate notch filter excludes the horizon emission within the region outlined by the orange dashed box.}
\label{fig:frf_sources}
\end{figure*}

\subsection{Fringe-rate Filtering}
\label{sec:fr-filter}

For interferometric drift-scan observations, which do not track any particular field on the sky and simply observe the sky as it passes through the baseline fringes, there is a useful mapping from time-based Fourier modes in the visibility (i.e., fringe-rates) to the location on the sky where the signal originated from.
\citet{Parsons2009} first studied this relationship for the purposes of low-frequency interferometric calibration.
This was later studied as a means for understanding both foregrounds and instrumental features seen in drift-scan observations \citep[e.g.][]{Shaw2014, Parsons2016}.
For HERA, these kinds of filters have been used to effectively mitigate instrumental systematics \citep{Kern2019, Kern2020b, Kern2022, Josaitis2022}.

Recently, \citet{Charles2023} proposed using fringe-rate filters to improve the fidelity of gain calibration for low-frequency 21\,cm interferometers. Their analysis focused solely on mitigating the impact of poorly modeled diffuse emission on absolute calibration, while Charles et al. (2023 in prep) investigates how this technique can mitigate nonredundancies arising from mutual coupling. Here, we adopt this fringe-rate filtering technique to study its efficacy in improving redundant calibration in the presence of nonredundancies caused by antenna beam variations. Specifically, we use a ``fringe-rate notch'' filter that rejects $f\sim0$ mHz modes in the data similar to the one described in \citet{Charles2023}.
In our paper~I, we showed that diffuse emission from the observer's horizon, which has $f\sim0$ mHz, is one of the main sources of the nonredundancy error, especially for the feed tilting offset (see Figure~7 of Paper~I). This implies that applying the fringe-rate notch filter should reduce nonredundancies in the data, and improve the calibration process.

Figure~\ref{fig:frf_sources} describes where different kinds of sky sources are located in the delay and fringe-rate space. We considered raw visibilities (i.e., no gains and calibration applied) for GLEAM (left), GSM (middle), and EoR (right) signals, separately to present Fourier transform as functions of the frequency and time. Two short baselines were chosen: the east-west direction (top panels) and the north-south direction (bottom panels). If we look at the delay axis first, most power of the GLEAM sources resides in the low delay modes spanning $-200$ to $200$~ns since the point sources are spectrally smooth. In the case of diffuse emission based on the GSM, there are additional features such as two central bright blobs from the pitchfork effect \citep{Thyagarajan2015a, Thyagarajan2015b} and extended power in high delay modes. As for the EoR signal, the spectrum is fluctuating along frequency, and thus the power is uniformly distributed in the delay domain.

\begin{figure*}[t!]
\centering
\includegraphics[scale=0.55]{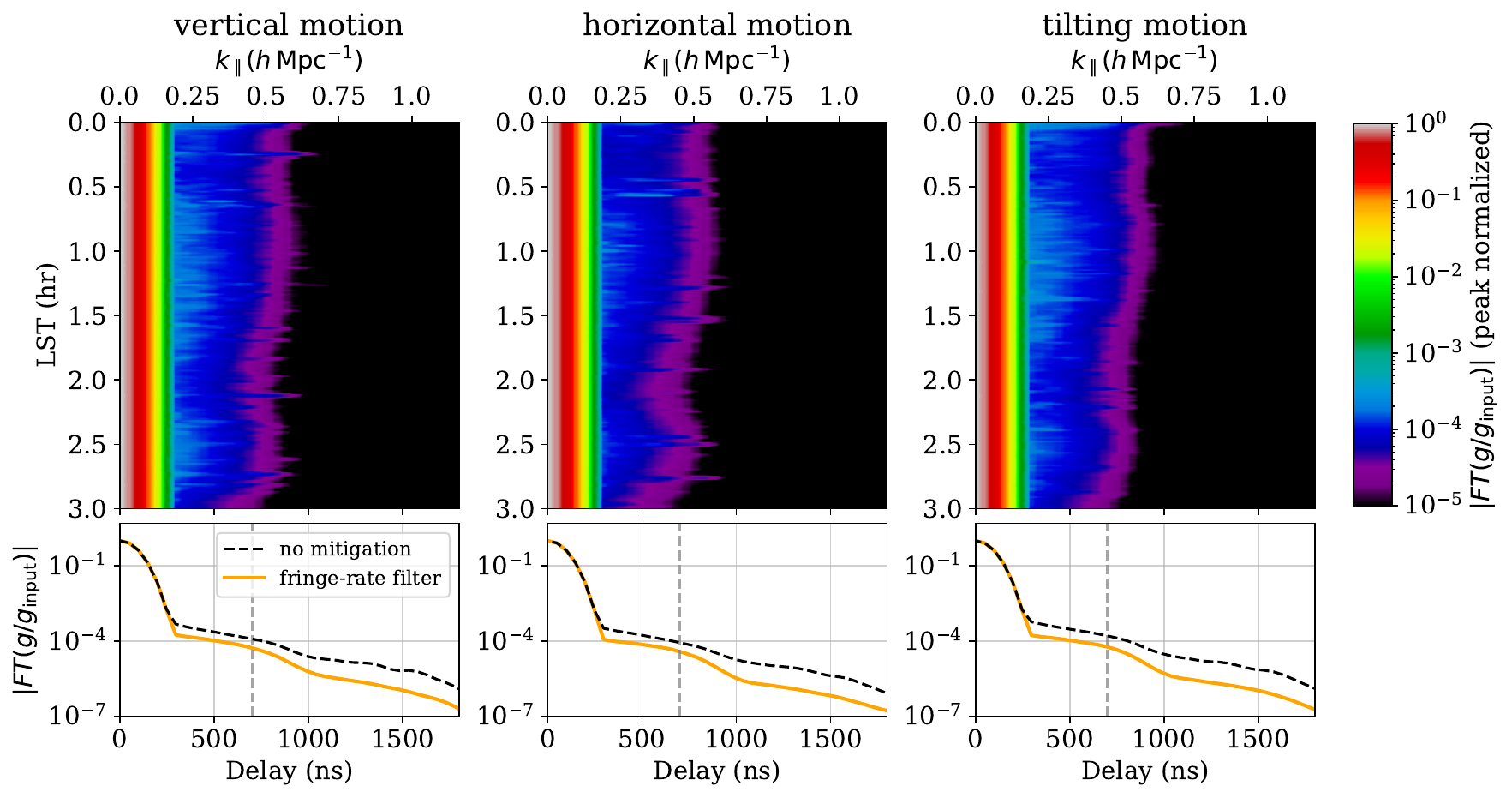}
\caption{Delay spectrum of the gain solutions with fringe-rate filtering. The format is the same as Figure~\ref{fig:fft_gain_errors_bl_cut_LST}. We considered the half filter width $f_w = 0.3$~mHz corresponding to the orange dashed box in Figure~\ref{fig:frf_sources}. The mitigation takes effect across all delays. The improvement at around 300--500~ns is notable compared to previous mitigation methods.}
\label{fig:fft_gain_errors_frf_LST}
\end{figure*}

Now let's consider the fringe-rate axis. The north-south baseline maps the sky inside the main lobe to near zero fringe rates and thus contains majority of the power at $f\sim0$~mHz as shown in the bottom panels. For the east-west baseline, sky sources in the main lobe of the primary beam, especially at zenith, have positive fringe rates and the main lobe drifts in the positive direction as the east-west baseline length increases. However, for the GSM, there is still a significant fraction of power remaining at zero fringe rate that originates from the horizon galactic emission. Namely, the fringe-rate notch filtering can suppress nonredundancies dominated by the diffuse horizon emission by removing signals at around zero fringe rates within the orange dashed box. Further details about the baseline-dependent relation between the sky positions and fringe rates are described in \citet{Parsons2016}.

For fringe-rate notch filtering (analogous to a high-pass temporal filtering), we adopted the \texttt{DAYENU}\footnote{The full name is DPSS Approximate lazY filtEriNg of foregroUnds. The software library is publicly available in \url{https://github.com/HERA-Team/uvtools}.} filter \citep{Ewall-Wice2020} which is a linear filter defined in a basis of Discrete Prolate Spheroidal Sequences \citep[DPSS;][]{Slepian1978}. This filter takes an analytic covariance model to suppress signal inside a half-filter width ($f_w$) with a suppression factor ($\epsilon$), where $\epsilon = 10^{-8}$ was chosen for this study. 

The high-pass DPSS-based filter is applied to the observed visibilities before the calibration along the time axis. The filtered visibilities are then calibrated through the redundant and absolute calibration steps with the model visibilities that are also fringe-rate filtered. An optimal half filter width of $f_w = 0.3$~mHz was determined by accessing degrees of mitigation on chromatic gain errors with various filter sizes. A smaller or larger than the optimal filter size returns a larger calibration bias.

Figure~\ref{fig:fft_gain_errors_frf_LST} summarizes results of the fringe-rate filtering mitigation. The strong bulge observed at around LST of 1~hr in Figure~\ref{fig:fft_gain_errors_no_mitigation_LST} has almost vanished, which indicates the effectiveness of the fringe-rate filter in suppressing the emission from the galactic plane lying on the horizon. Compared to the baseline cut-off mitigation, the frequency structure at $\sim$700~ns is less effectively reduced. However, the delay spectrum at low delay modes, $\sim$300--500~ns, is a few factor lower than the case without any mitigation, which is not accomplished by other mitigation methods.

As different mitigation techniques result in different mitigated gain results, we anticipate combining the mitigation methods would improve the overall gain solution, as discussed in the following section.

\begin{figure}[t!]
\centering
\includegraphics[scale=0.55]{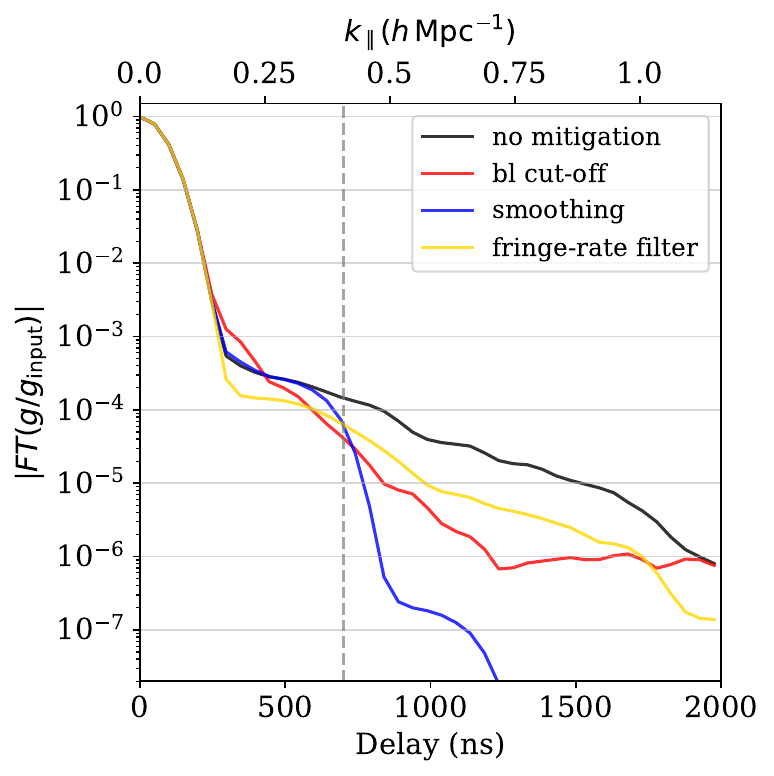}
\caption{Delay spectra with the three gain error mitigation techniques are contrasted to the delay spectrum before mitigation is applied (black curve). Without loss of generality, only the results for vertical feed perturbations are shown. The delay spectra at LST~=~1~hr are averaged over unflagged antennas. All different strategies provide mitigated delay spectrum results but in different shapes. See the main context for more details.}
\label{fig:fft_gain_summary}
\end{figure}

\begin{figure}[t!]
\centering
\includegraphics[scale=0.55]{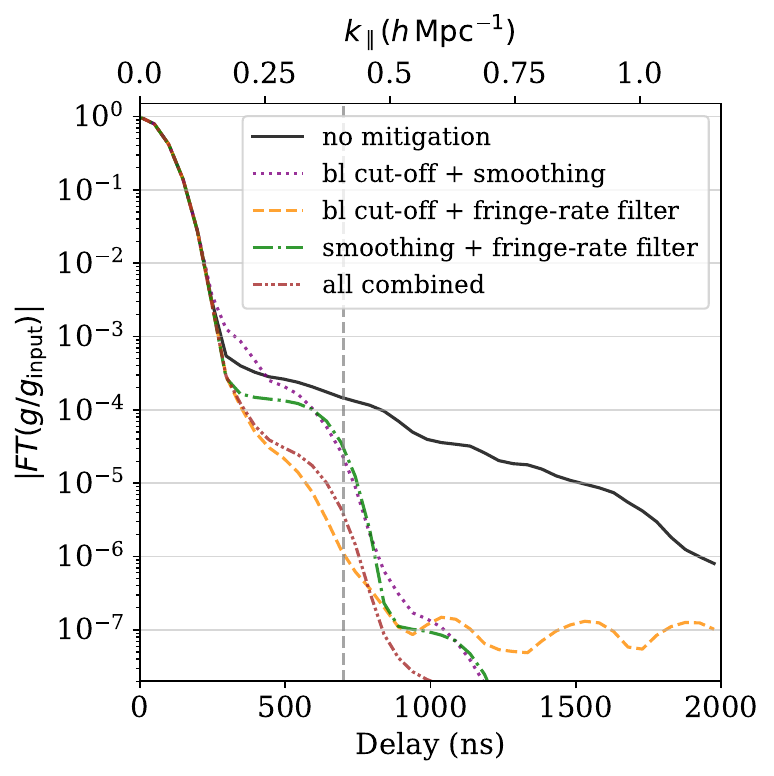}
\caption{Delay spectra with different combinations of mitigation strategies. The delay spectra are shown for the vertical feed motion with a sky at LST~=~1~hr. The most effective method (orange curve) is the combination of baseline cut-off and fringe-rate filtering. It brings the suppression to $10^{-6}$ at $\kpara \sim 0.4 \, \hMpcinv$ to satisfy the EoR detection requirements.}
\label{fig:fft_gain_combined}
\end{figure}

\begin{figure*}[t!]
\centering
\includegraphics[scale=0.55]{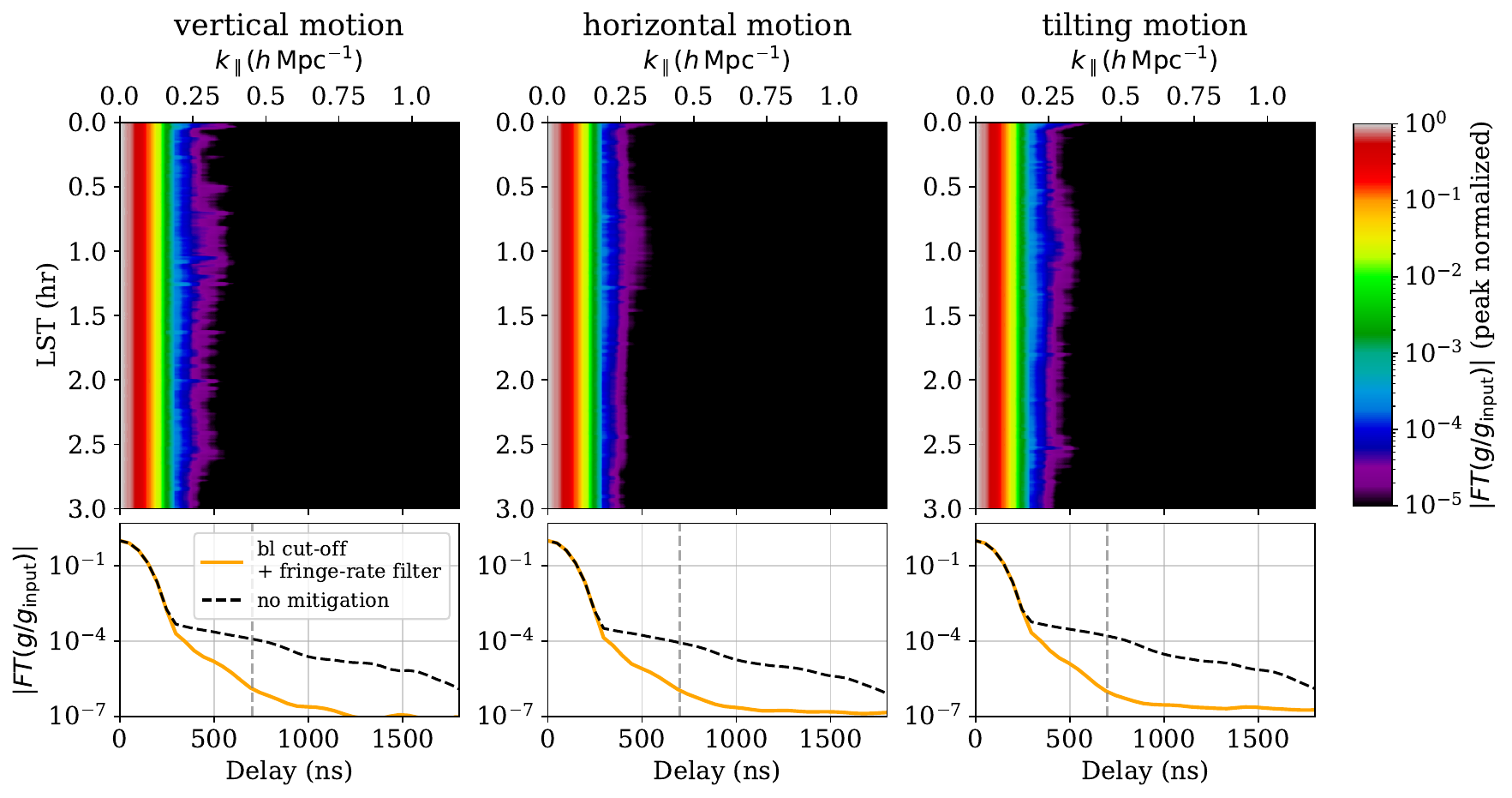}
\caption{Delay spectrum of the gain solutions with both baseline cut-off and fringe-rate filtering mitigation applied. The leakage due to chromatic gain errors shown in Figure~\ref{fig:fft_gain_errors_no_mitigation_LST} retreats to $\kpara \sim 0.3 \, \hMpcinv$ over all LSTs, which indicates that gain errors are suppressed enough to recover the EoR window on the power spectrum.}
\label{fig:fft_gain_errors_bl_cut_frf_LST}
\end{figure*}

\subsection{Summary of Mitigation}
In this section, we summarize the three mitigation methods and discuss their hybrid applications. Figure~\ref{fig:fft_gain_summary} characterizes the result of each mitigation at LST~=~1~hr, showing the delay spectra with suppressed power at high delay modes. Without loss of generality, only the vertical feed perturbation cases are represented, as all feed motions share a similar trend. All mitigation techniques alleviate the chromatic gain errors but in different ways. The baseline cut-off mitigation that down-weights long baselines ($> 87.6$~m) in calibration mainly reduces the fine-frequency structure induced by the baselines, and thus suppresses the delay spectrum at high delays. Meanwhile, the fringe-rate notch filtering, excluding diffuse galactic emission located at nearly zero fringe rates across all delays to 2000~ns, drops the delay spectrum by at least a few factors beyond 300~ns. The smoothing that rules out fine spectral structure with the low-pass frequency filter brings a rapid fall of the delay spectrum above 700~ns, but has little improvement below 500~ns.

Since different mitigation approaches alleviate the chromatic gain errors in different ways, a combination of the mitigation techniques can provide a more effective way to control the leakage. In Figure~\ref{fig:fft_gain_combined}, four different combinations of the methods are presented: baseline cut-off + smoothing (purple), baseline cut-off + fringe-rate filtering (orange), smoothing + fringe-rate filtering (green), and all three combined method (brown). Either the baseline cut-off or the fringe-rate filtering method together with smoothing results in a fast drop at delays $> 700$~ns, but there is little improvement at delays $< 500$~ns compared to the method alone.

The combination of baseline cut-off and fringe-rate filtering is distinct from the smoothing joint cases. Limiting the baselines to short ones leads to less nonredundancy errors from point sources and applying the fringe-rate notch filter reduces the nonredundancy caused by the diffuse emission. As a result, this combination accounts for the nonredundancy effects for both types of sources, and substantially suppresses the spectral artifacts arising from calibration. The orange line in Figure~\ref{fig:fft_gain_combined}, with the baseline cut-off~=~87.6~m and $f_w$~=~0.3~mHz for the fringe-rate filter, shows that the delay spectrum is significantly reduced across all delay modes, and can reach down to $10^{-6}$ at $\kpara \sim 0.4 \, \hMpcinv$, which in return helps recover the clean EoR window similarly to the fiducial one. The fully combined case also reduces the delay spectrum considerably but there is a bump artifact created by the low-pass filter for the smoothing at 500~ns.

Figure~\ref{fig:fft_gain_errors_bl_cut_frf_LST} presents the delay spectrum along LST when the baseline cut-off and fringe-rate notch filtering are both applied. Across all LSTs, the bleeding gain errors shown in Figure~\ref{fig:fft_gain_errors_no_mitigation_LST} are reduced significantly, restrained to $\kpara \sim 0.3 \, \hMpcinv$ for all three types of feed perturbations, which offers at least two orders of magnitude in improvement at $\kpara \sim 0.4 \, \hMpcinv$.

So far, we have mainly focused on the mitigation effect in terms of the gain solutions. The gain solutions will be applied to the observed visibility, and calibration errors will propagate to the calibrated visibility. One way to assess the calibration errors imprinted in the visibility is to compute the power spectrum. In addition, our test is limited to a certain size of feed perturbations: $\sigma_{\rm feed} = 2$~cm for axial and lateral feed motions and $\sigma_{\rm feed} = 2^\circ$ for feed tilting. For a comprehensive understanding on the effectiveness of the mitigation, we test various sizes of feed perturbations and discuss the mitigation effect on the power spectrum in Section~\ref{sec:pspec}.

\section{Power Spectrum Estimation}
\label{sec:pspec}

In Section~\ref{sec:redcal} and \ref{sec:mitigation},  we introduced the delay spectrum of gain solutions. Equivalently, the visibility delay spectrum can be produced by performing the frequency Fourier transform on the calibrated visibilities,
\begin{equation}
    \Tilde{V}(\textbf{u}, \tau) = \int d\nu\, w(\nu) V(\textbf{u}, \nu) e^{2\pi i \nu \tau},
    \label{eqn:delay_spectrum}
\end{equation}
where $\textbf{u} = \textbf{b}/\lambda$ and $w(\nu)$ is the seven-term Blackman-Harris tapering window function. According to Equation~\eqref{eqn:delay_spectrum}, for a given baseline, the smooth frequency spectrum of the foreground confines its delay spectrum to near zero delay modes with a width limited by the horizon limit, $\tau_{\rm hor} = |\mathbf{b}|/c$ \citep{Datta2010, Vedantham2012, Morales2012, Morales2018, Parsons2012, Trott2012, Thyagarajan2013, Liu2014, Pober2014}. In contrast, the 21-cm cosmological signal with fluctuating spectral structure allows its delay spectrum to spread beyond the horizon limit.

\subsection{2D Power Spectrum Estimation}
\begin{figure}[t!]
\centering
\includegraphics[scale=0.40]{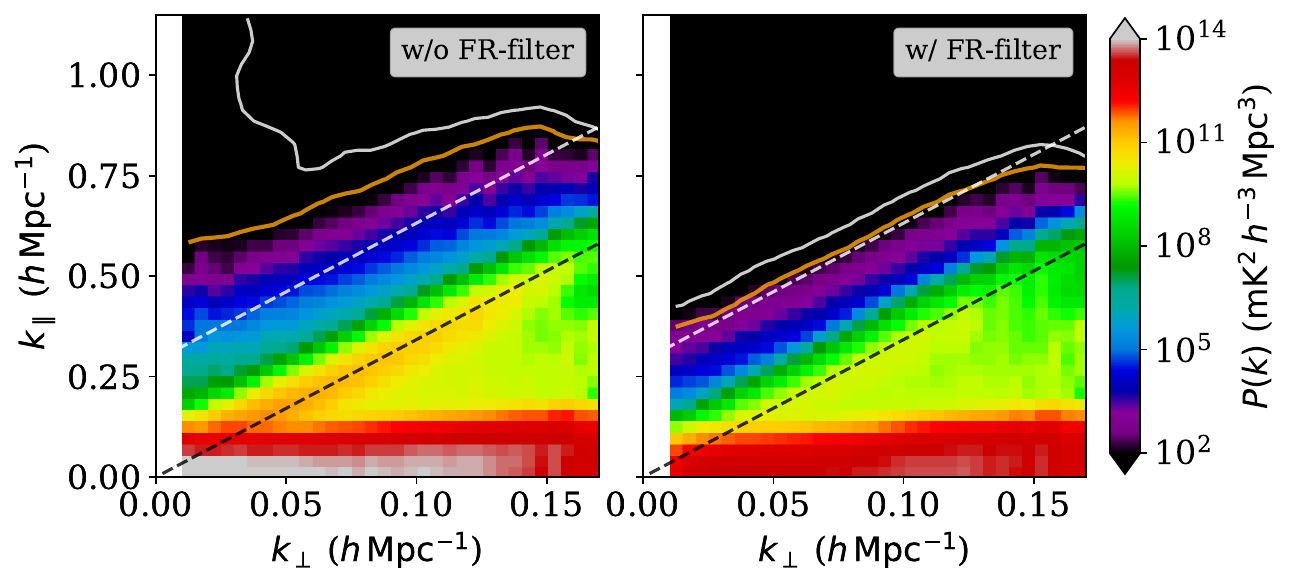}
\caption{Foreground power spectrum estimation with the fiducial beam model before (left) and after (right) the fringe-rate notch filter. Orange and white contours represent the EoR power equal to and 5 times larger than the foreground power, respectively. The pitchfork along the horizon limit (black dashed line) shown in the left panel is suppressed in the right panel after the fringe-rate notch filtering. As a result, the spillover of the foregrounds is constrained inside the buffer line (white dashed line), offering a room for the EoR detection above the white dashed line. Note that the fringe-rate filter here is not being applied to visibilities prior to calibration but applied to visibilities after calibration prior to forming the power spectrum merely to suppress the pitchfork.}
\label{fig:pspec_fiducial}
\end{figure}

The 2D power spectrum, or cylindrically averaged power spectrum, is a tool to separate the cosmological signals from the foregrounds. It is defined by cosmological Fourier modes perpendicular and parallel to the line-of-sight, $\kperp$ and $\kpara$, which are linearly scaled from $|\mathbf{u}|$ and $\tau$. More specifically, the relationship is
\begin{equation}
    \kperp = \frac{2\pi|\textbf{u}|}{X}, \, \kpara = \frac{2\pi\tau}{Y},
\end{equation}
where $X = D(z)$, $Y = c(1+z)^2/(\nu_{0}H(z))$, $D(z)$ is the comoving distance, and $H(z)$ is the Hubble parameter. With increasing $\kperp$, $k_{\rm \parallel, hor}$ corresponding to $\tau_{\rm hor}$ also increases, forming the foreground wedge below the horizon limit $k_{\rm \parallel, hor} = \frac{H(z)D(z)}{c(1+z)} \kperp$.

The 2D power spectrum estimate is calculated from the delay spectrum as follows,
\begin{equation}
    \hat{P}(\kperp, \kpara) = \frac{X^2Y}{B_{\rm pp}\Omega_{\rm pp}} |\Tilde{V}(\textbf{u}, \tau)|^2,
    \label{eqn:pspec}
\end{equation}
where $B_{\rm pp} = \int d\nu\, |w(\nu)|^2$ and $\Omega_{\rm pp}$ is the spatial integral of the squared primary beam \citep{Parsons2014}. The complex visibility delay spectra were averaged over redundant baselines and fed into Equation~\eqref{eqn:pspec}. We implemented the power spectrum calculation via the publicly available software, \texttt{hera\_pspec}\footnote{\url{https://github.com/HERA-Team/hera_pspec}}.

Figure~\ref{fig:pspec_fiducial} shows the 2D foreground power spectrum with the fiducial beam model. The power spectrum is averaged over LST of 0--3~hr. The foregrounds are indicated in the color bar. We also over-plotted the cosmological signal in contours, the orange and white solid contours corresponding to the region that the EoR power is equal to and 5 times larger than the foregrounds, respectively. The black dashed line indicates the horizon limit, and the white dashed line is for a 500-ns buffer above the horizon limit.

Because the fiducial case does not suffer from calibration errors, in principle the foreground should not leak into high $\kpara$ modes beyond the horizon limit. In the left panel, we see the foreground power still spills into the EoR window above the horizon limit, which is different from the leakage due to calibration errors. This is because of the pitchfork effect, along the horizon limit line, that is convolved with the primary beam. The pitchfork and associated spillover can be reduced by the fringe-rate notch filter \citep{Thyagarajan2016, Kern2020a, Kern2020b} and thus we applied the fringe-rate notch filter with $f_w = 0.3$~mHz to calibrated visibilities.

In addition, as illustrated in Figure~\ref{fig:frf_sources}, the EoR signals with short east-west baselines are influenced by a significant signal loss. Therefore, in forming the power spectrum we included only baselines longer than 15~m in the east-west direction for both foreground and EoR power spectra. See Kern et al. 2021 (HERA memo \#96\footnote{\url{http://reionization.org/science/memos/}}) for more discussion on the signal loss due to the fringe-rate notch filter. In the right panel of Figure~\ref{fig:pspec_fiducial}, after the fringe-rate filtering, the bright foreground along the black dashed line, the pitchfork, is suppressed and as a result, the intrinsic spillover above the white dashed line is also reduced. In the following section, the power spectra with perturbed beams are also results of the fringe-rate filtering to suppress the pitchfork.

\begin{figure*}[t!]
\centering
\includegraphics[scale=0.5]{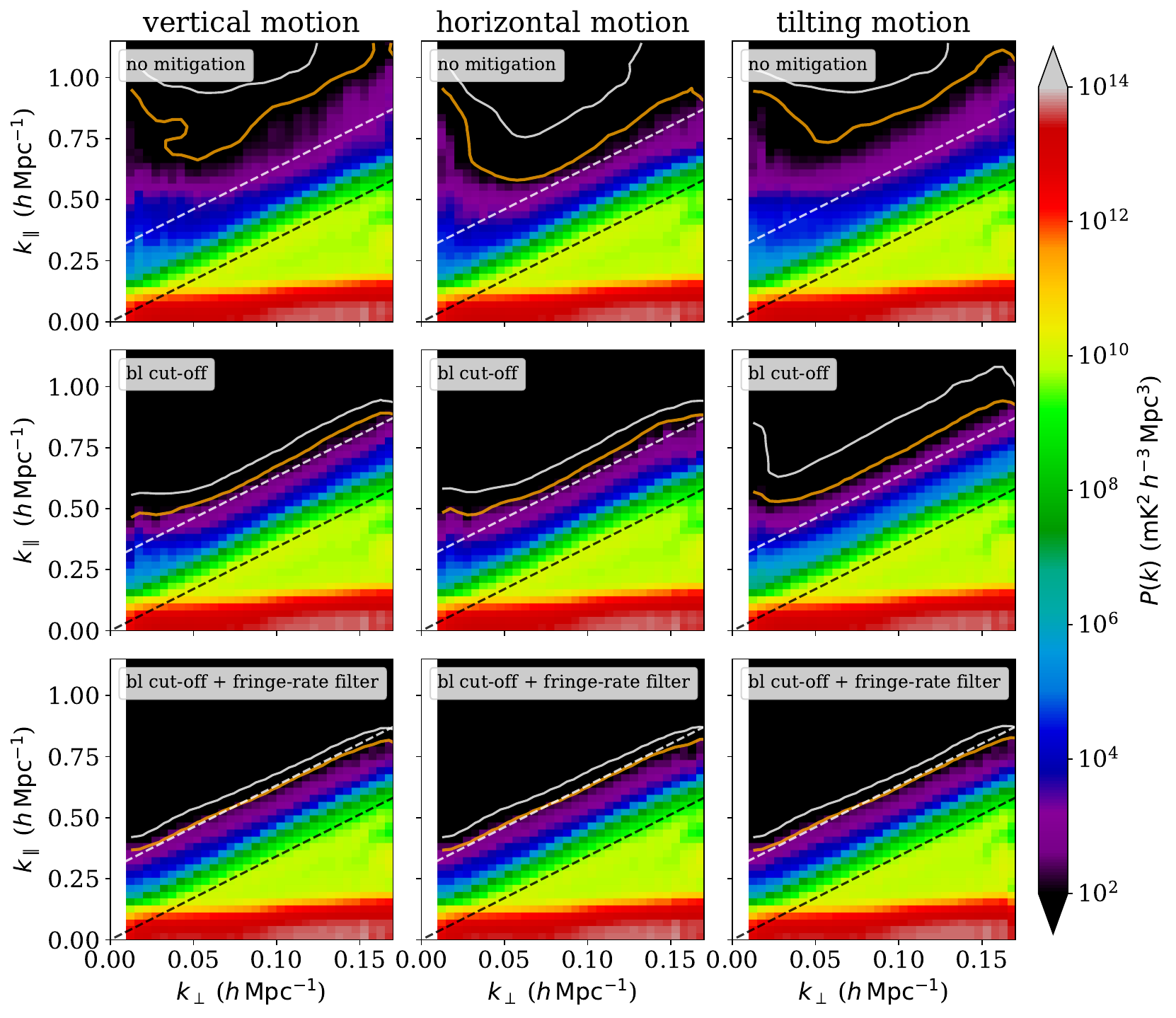}
\caption{Foreground power spectrum estimation for perturbed beams without and with the mitigation strategies. Top row shows the foreground power spectrum for the vertical (left column), horizontal (middle column) and tilting (right column) motions before the mitigation on the chromatic gain errors. The foreground leakage contaminates the cosmological modes that are available for the EoR detection above the buffer line (white dashed line). The middle row presents the results after the baseline cut-off mitigation with 87.6~m cut-off. The mitigation reduces the leakage to some extent, but the low Fourier modes are still corrupted by the foreground contamination. In the bottom row, the combined method of the baseline cut-off and fringe-rate filtering leads to a considerable reduction of the foreground leakage for all types of feed perturbations and recovers the clean EoR window similar to the fiducial power spectrum shown in Figure~\ref{fig:pspec_fiducial}.}
\label{fig:2d_pspec_mitigated}
\end{figure*}

\subsection{Power Spectrum Estimation with Mitigation}
As discussed above, in the ideal case when there is no calibration bias, the foreground is confined to the foreground wedge. However, when there are chromatic gain errors due to per-antenna perturbations, the convolution of gain errors with the observed visibility in Fourier space can result in foreground leakage on the power spectrum, analogous to Figure~\ref{fig:fft_gain_errors_no_mitigation_LST}. In the top panels of Figure~\ref{fig:2d_pspec_mitigated}, the resultant power spectra without applying any mitigation strategy, averaged across 0--3~hr, are shown. All feed motions display the corrupted EoR window with the detection border line (orange contour) pushed beyond $\kpara \sim 0.7\,\hMpcinv$, losing potential detectability at low $k$ modes where relatively higher sensitivity can be obtained when the thermal noise is included. The foreground leakage is predominantly derived by the 0.5--1.5~hr gain corruption shown in Figure~\ref{fig:fft_gain_errors_no_mitigation_LST}.

The middle panels show the case for the mitigation with baseline cut-off that has been explored by previous studies \citep{Ewall-Wice2016, Orosz2019}. We set 87.6~m cut-off for the mitigation. The foreground corruption in the EoR window is largely restrained compared to no mitigation case, which is consistent with the finding of \citet{Orosz2019}. One notable thing is that the low $\kperp$ modes are still corrupted by the foreground leakage, making the EoR detection difficult at low $k$ modes without further suppression of the foreground contamination. For the tilting motion, the mitigation is less effective than other motions as the gain errors due to the tip-tilt primarily derive from diffuse emission.

With the combined strategy of the baseline cut-off and fringe-rate filtering, we showed that chromatic gain errors are significantly improved. In the bottom panels of Figure~\ref{fig:2d_pspec_mitigated}, the results with the mitigated gains are given. Regardless of the feed motions, the mitigation method removes the foreground contamination and achieve as clean a EoR window as the fiducial model, enabling EoR detectability at $\kpara = 0.3$--$0.5\,\hMpcinv$ where the cosmological signal detection may be feasible with a high sensitivity.

\begin{figure*}[t!]
\centering
\includegraphics[scale=0.55]{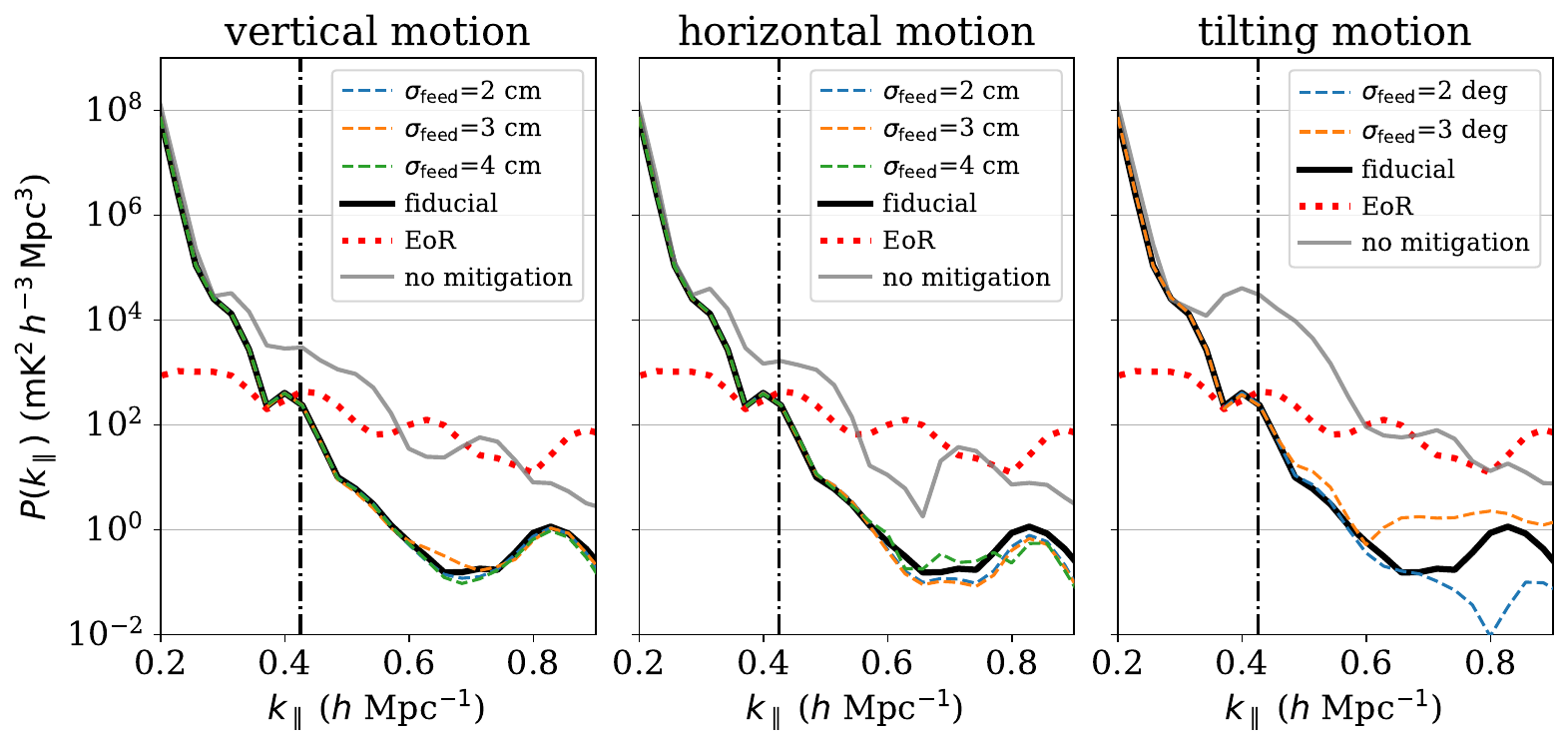}
\caption{1D power spectra along $\kpara$ at $\kperp = 0.04\,\hMpcinv$ for the foregrounds (dashed and solid lines) and the EoR (red dotted line). The vertical line indicates the 500-ns buffer line. We present the case without mitigation for $\sigma_{\rm feed}$ = 2~cm or 2$^\circ$ (gray solid line) that corrupts $k$ modes in the EoR window by the foreground residual. The combined mitigation method with the baseline cut-off and fringe-rate filtering makes the perturbed power spectra comparable to the fiducial one, and leave the EoR detection feasible beyond the buffer line, regardless of the sizes of perturbations.}
\label{fig:1d_pspec_mitigated}
\end{figure*}

Paper~I shows that larger feed motions cause larger calibration errors, and thus larger amounts of foreground leakage. With the mitigation technique, we examined how the foreground contamination can be alleviated for the larger perturbation cases. Figure~\ref{fig:1d_pspec_mitigated} presents 1D power spectra along $\kpara$ at $\kperp = 0.04\,\hMpcinv$, averaged over 0--3~hr. From left to right, results for vertical, horizontal, and tilting motions are given. For the vertical and horizontal feed motions, $\sigma_{\rm feed}$ = 2, 3, and 4~cm models are considered and for the tip-tilt, $\sigma_{\rm feed}$ = 2$^\circ$ and 3$^\circ$ models are used. For comparison, we also show the case without mitigation with $\sigma_{\rm feed}$ = 2~cm and 2$^\circ$ for the translation and tilting motion, respectively. The vertical line indicates the 500-ns buffer, $k_{\rm \parallel, buffer} = k_{\rm \parallel, hor} + 0.29~\hMpcinv$.

The foreground power spectrum with no mitigation (gray solid line) shrinks available $k$ modes for the EoR detection compared to the fiducial case (black solid line). We found that the joint mitigation method with the baseline cut-off and the fringe-rate filtering (dashed lines) makes the foreground power in the EoR window comparable to the fiducial model, regardless of the types of feed motions and the perturbation sizes. The excess EoR power spectrum beyond the buffer line relative to the foregrounds means the EoR detection is free from the foreground bias. However, we noticed that as the perturbation size increases, the calibration pipeline tends to fail to converge to the global minimum because of large nonredundancies, ending up flagging more data to keep well-calibrated results as shown in Figure~\ref{fig:1d_pspec_mitigated}. The models with $\sigma_{\rm feed} >$ 2~cm or 2$^\circ$ result in the unstable calibration solutions, which may indicate the feed motion should be constrained to $\sigma_{\rm feed} =$ 2~cm or 2$^\circ$ perturbations for better calibration performance.

\section{Conclusions}
\label{sec:conclusions}
Feed motions cause a deviation of the primary beam shape from the fiducial one. The per-antenna perturbations from the feed motions cause nonredundancies in measured visibilities and introduce artificial spectral structure in the calibration gain solutions from the redundant and absolute calibration, employed by HERA, propagating to the power spectrum estimation. The perturbation in the beam shape and the resultant chromatic gain errors are discussed in Paper~I.

To extend the work of Paper~I, we simulated longer LST mock observations covering 0--3~hr including the GLEAM sky model and the GSM that represent point sources and diffuse emission, respectively. Analogous to Paper~I, we separated the feed motions into the vertical, horizontal, and tilting motions. All feed motions introduce artificial structure in the calibration gains compared to the fiducial model, but with different overall aspects for different feed motions as shown in Figure~\ref{fig:gain_errors_with_LST}. It can be understood in terms of the different behaviors of the primary beam induced by each feed motion. In addition, fine spectral and temporal gain structure is produced for all feed motions. The gain delay spectrum shows that the chromatic gain errors corrupt the delay modes $\kpara > 0.2 \, \hMpcinv$, forming a bulge at LST~=~0.5--1.5~hr where the galactic plane are located near the horizon (Figure~\ref{fig:fft_gain_errors_no_mitigation_LST}). Mitigating the chromatic gain errors is the main goal of this study: attaining $\sim$$10^{-5}$ suppression of the gain delay spectrum at $\leq 700$~ns (or $\kpara \leq 0.4 \,  \hMpcinv$).

We have explored three different mitigation techniques: placing a limit on the maximum baseline length used for the calibration, smoothing the gain structure along the frequency axis after the calibration, and applying the fringe-rate notch filter on the measured visibilities prior to the calibration. All mitigation methods reduce the fine spectral artifacts to some extent but the aspects of the improvement are different. The baseline cut-off method constrains the nonredundancy arising from long baselines and results in a reduction in the fine-frequency artifacts at $> 400$~ns. The gain delay spectrum is suppressed to $\sim$$5 \times 10^{-5}$ at 700~ns, which is still a bit higher than desired. The smoothing with the low-pass frequency filter effectively smooths out the small scale frequency features, presenting the gain delay spectrum with a rapid drop after $\sim$700~ns. Nonetheless, it has little effect at delays $< 500$~ns. The fringe-rate notch filtering improves the calibration process by eliminating the diffuse horizon emission, one of the main nonredundancy sources, from the visibility measurements. It leads to the delay spectrum mitigated by a few factor across all delay modes $> 300$~ns, which is still not sufficient to acheive the required suppression.

Because each method alone is not enough to accomplish the required suppression of the chromatic gain errors, joint mitigation strategies are also examined. The combinations between the smoothing and other methods only improve the delay spectrum beyond 700~ns and low delay modes have little improvement. When the baseline cut-off method is joint with the fringe-rate filtering, there is substantial improvement. Because the baseline cut-off and the fringe-rate filtering strategies are mainly responsible for reducing the nonredundancies from the point sources and diffuse horizon emission, respectively, the joint method addresses both nonredundancy effects. It brings the delay spectrum down to $\sim$$10^{-6}$ at 700~ns that satisfies the need for suppression of gain errors for EoR detection.

As the chromatic gain errors propagate to the power spectrum estimate, we see that the EoR window is corrupted by the foreground leakage in the 2D power spectra when there is no mitigation for all feed motions. If the baseline cut-off mitigation is applied as \citet{Ewall-Wice2016} and \cite{Orosz2019} did, the foreground contamination is suppressed to some degree but some contamination still remains in the low $k$ modes. The combined method of the baseline cut-off and the fringe-rate filtering clears away the contamination in the EoR window and arrives at a power spectrum similar to the fiducial one with perfect calibration.

In Section~\ref{sec:redcal}--\ref{sec:pspec}, we focused on the modest feed motions by $\sigma_{\rm feed} = 2$~cm and $2^\circ$ for the translation motions and the tip-tilts, respectively. We also investigated the effectiveness of the mitigation technique with larger perturbation sizes. We found that the 1D power spectra with the combined mitigation of the baseline cut-off and fringe-rate filtering present the results similar to the fiducial one even for $\sigma_{\rm feed} = 4$~cm or $3^\circ$, demonstrating that the mitigation technique is effective enough to suppress the chromatic gain errors for larger perturbations. However, we also noticed that the larger nonredundancies due to the feed motion may lead to failure in finding appropriate gain solutions, which implies we need to constrain the feed motions to $\sigma_{\rm feed} = 2$~cm and $2^\circ$.

In this study, we only focused on the middle band corresponding to the redshift of $\sim$7. HERA is also designed to study the Cosmic Dawn at lower frequencies. The low band with a wider primary beam may bring about a larger per-antenna perturbation than the middle band for the same feed position deviations. Hence, it is important to test the mitigation scheme for the low band as well to set a requirement for feed positioning with a more comprehensive analysis. We defer this to future work.

\section*{Acknowledgements}
This material is based upon work supported by the National Science Foundation under Grant Nos. 1636646 and 1836019 and institutional support from the HERA collaboration partners.  This research is funded in part by the Gordon and Betty Moore Foundation through Grant GBMF5212 to the Massachusetts Institute of Technology.
N. K. acknowledges support from the MIT Pappalardo fellowship. The National Radio Astronomy Observatory is a facility of the National Science Foundation operated under cooperative agreement by Associated Universities, Inc. B. D.is a Jansky Fellow of the National Radio Astronomy Observatory.

\bibliography{main}{}
\bibliographystyle{aasjournal}

\end{document}